\documentclass[a4paper,fleqn,usenatbib]{mnras}

\usepackage{newtxtext,newtxmath}

\usepackage[T1]{fontenc}
\usepackage{ae,aecompl}

\usepackage{graphicx}	
\newcommand{\xten}[1]{\mbox{$\times 10^{#1}$}}

\newcommand{\ltappeq}{\raisebox{-0.6ex}{$\,\stackrel
{\raisebox{-.2ex}{$\textstyle <$}}{\sim}\,$}}
\newcommand{\gtappeq}{\raisebox{-0.6ex}{$\,\stackrel
{\raisebox{-.2ex}{$\textstyle >$}}{\sim}\,$}}

\title[Cosmic ray induced desorption of ices]
{Towards a better understanding of ice mantle desorption 
by cosmic rays}

\author[J. M. C. Rawlings]
{Jonathan M. C. Rawlings,$^{1}$\thanks{E-mail: jcr@star.ucl.ac.uk}\\
\\
$^{1}$Department of Physics and Astronomy, University College London,
Gower Street, London, WC1E 6BT, UK}

\date{....}

\pubyear{2022}

\begin{document}
\label{firstpage}
\pagerange{\pageref{firstpage}--\pageref{lastpage}}
\maketitle

\begin{abstract}
The standard model of cosmic ray heating-induced desorption of 
interstellar ices is based on a continuous representation of
the sporadic desorption of ice mantle components from classical 
($0.1\mu$m) dust grains.
This has been re-evaluated and developed to include tracking the 
desorption through (extended) grain cooling profiles, consideration 
of grain size-dependencies and constraints to the efficiencies.
A model was then constructed to study the true, sporadic, nature of the 
process with possible allowances from species co-desorption and whole
mantle desorption from very small grains.
The key results from the study are that the desorption rates are highly 
uncertain, but almost certainly significantly larger than have been 
previously determined.
For typical interstellar grain size distributions it is found that
the desorption is dominated by the contributions from the smallest 
grains. 
The sporadic desorption model shows that, if the interval between 
cosmic ray impacts is comparable to, or less than, the freeze-out timescale,
the continuous representation is inapplicable;
chemical changes may occur on very long timescales, resulting in strong 
gas phase chemical enrichments that have very non-linear dependences on the 
cosmic ray flux.
The inclusion of even limited
levels of species co-desorption and/or the contribution from very small
grains further enhances the rates, especially for species such as H$_2$O.
In general we find that cosmic-ray heating is
the dominant desorption mechanism in dark environments. 
These results may have important chemical implications for protostellar 
and protoplanetary environments.
\end{abstract}


\section{Introduction}
\label{sec:intro}

In the dense, cold, environments of molecular clouds the, often very
rich, chemistry is determined by both gas-phase and solid-state
reactions. These are interconnected by the freeze-out of gas-phase 
species and the sublimation, or desorption, of solid-state species.

In dense gas with a cold dust component, freeze-out is very rapid 
(and typically faster than the timescale for dynamical evolution), so that 
the desorption processes are required to explain the observed abundances
of gas-phase species \citep[e.g.][]{TMCW04,Cas12}.
Indeed, the observed presence of tightly bound complex organic molecules 
\citep[e.g.][]{Cern12} is 
indicative of the presence not just of efficient desorption, but also of
energetic processes that drive the chemistry of their formation.

Obviously, to understand the chemical composition and evolution of
a molecular cloud or star-forming region, it is essential to have an
accurate understanding of the desorption processes.
A variety of mechanisms have been proposed but, broadly speaking, 
these can be divided into continuous processes, and 
sporadic processes \citep{RRVW07}. Examples of the former include thermal
desorption \citep{Wat76}, photodesorption (direct and cosmic-ray induced) - often 
driven by the photodissociation of molecules in the surface layers of ice 
mantles, and desorption that is driven by the 
enthalpy of molecule formation in the solid-state \citep[e.g.][]{Min16}.
Examples of the latter include ice mantle `explosions' following the
release of trapped chemical energy \citep[e.g.][]{CCP10} and desorption driven by the 
impact and passage of cosmic-ray particles through a grain 
\citep[e.g.][]{LJO85,HH93,BJ04,HC06}.

In recent years considerable attention has been paid to photodesorption
processes, which may operate dissociatively (e.g. for H$_2$O) or 
non-dissociatively (e.g. for CO) in the surface layers of ice mantles 
\citep[e.g.][]{Ob09,ODL09}. 
Such processes, which typically have a yield of $\sim 10^{-3}$, are 
believed to be particularly important in regions of intermediate extinction
($A_{\rm v}\sim 1-10$) due to direct photodesorption by the attenuated 
interstellar radiation field, but also in regions of higher extinction, where
the UV radiation field is dominated by the indirect cosmic ray excitation of 
H$_2$ Lyman and Werner bands \citep{PT83}. 
Indeed, photodesorption is often taken to be the dominant desorption process 
in cold, dense environments. It has been suggested that this may even 
be true in very high density, dark cores \citep{Cas12}.

Desorption can also be effected as a result of impulsive heating 
by (heavy ion) cosmic ray particles as they pass through grains.
However, whilst photodesorption has been very well studied, both 
theoretically and in the laboratory, the same cannot be said for  
cosmic-ray induced desorption and the effective rates that have 
been determined for this process are almost entirely theoretical.

In the standard representation, which we describe in the following section, 
the mechanism is one of sporadic heating by cosmic ray impact events, leading 
to grain heating and the rapid thermal desorption/sublimation of ice mantle 
species.
These impacts are rare (typically occuring once every Myr) but they
promote very efficient, possibly even catastrophic, desorption so that 
the net effect on the chemical abundances (particularly of gas-phase 
species) may be highly significant.

In this paper, we argue that - even within the commonly adopted paradigm 
in which cosmic-ray induced desorption is represented by a continuous 
process - following periodic whole grain heating/desorption by cosmic rays
of a single type and energy - the rates have been severely underestimated.
There are a variety of reasons for this; including an 
under-estimation of the grain cooling/ice evaporation timescale, a lack
of consideration of ice species co-desorption and the effects of 
consideration of varying grain sizes and cosmic ray properties/energies.
The last of these has been considered in recent publications (see Section 
\ref{sec:compare} below) and 
we do not discuss it in this study. However, the other issues have not been 
considered comprehensively, nor has the effect of simplifying the sporadic 
grain heating to a continuous desorption process.  
Once these factors have been included we find that, for most species of 
astrochemical interest, cosmic ray-induced desorption is almost certainly 
the dominant desorption mechanism in the denser, darker parts of 
star-forming regions, but remains very poorly constrained.

The structure of the paper is as follows:
In section \ref{sec:crdes} we give an outline of the physical processes that 
are involved in cosmic-ray induced desorption and
describe the continuous desorption representation that is widely adopted in
astrochemical models. The exension of this model to more realistic grain size
distributions is explained in section \ref{sec:MRN}, whilst in section 
\ref{sec:limits} we consider the limitations of that model, and their 
implications, as well as the possible major contribution to the process that
is made by very small grains.
In section \ref{sec:model} we describe a model that includes the sporadic nature
of the desorption events, as well as the possibility of ice species co-desorption,
and the results from that model are presented and 
discussed in section \ref{sec:results}.
Section \ref{sec:compare} compares our findings to other, recent, studies and 
a summary of our findings and the conclusions are given in section 
\ref{sec:summary}.

\section{Cosmic-ray induced desorption}
\label{sec:crdes}

The physics of the interaction between a cosmic ray particle 
(typically, a high energy massive ion) and an ice-coated grain is complex. 
In the context of how that interaction effects ice mantle sublimation or 
desorption the process is often (over-)simplified as being represented by 
the sporadic heating of the whole grain.
The resulting desorption that is driven by the heating of grains and the
sublimation of ices is very poorly understood/quantified.

In fact, the interaction between a cosmic ray and an icy grain has several 
physical stages \citep{BJ04}:
as the particle passes through a grain it causes excitations,
ionizations and deposits energy within an approximate cylinder
of interaction, with a typical radius of $\sim 100$\AA. The instantaneous 
effects of this interaction are initially manifest at the areas on the surface
of the grain where the cosmic ray particle enters and exits, the so-called
`hot spots', and spread to the rest of the grain as a result of thermal 
diffusion.
The nature and physics of this interaction is empirically 
ill-defined, but molecular dynamics simulations suggest that the physical 
processes can be divided into four phases:
(i) initial excitations and primary/secondary ionizations. 
It has been estimated that as much as 40\% of the impact energy could be 
carried away by fast electrons \citep{LJO85}, 
(ii) relaxation/recombinations, leading to prompt desorption of hot atoms 
from the `hot spots', possibly leading to a heat-spike pressure pulse capable of
significant energy transfer,
(iii) localised thermal diffusion and evaporation - `thermal spike 
sputtering' and then, finally, (iv) thermal diffusion throughout the grain -
`whole grain heating' (WGH) and evaporative cooling.
The relative importance of the different phases (and the resulting desorption 
of the ices) will depend on a number of factors.  
So, for example, it is obvious that if the grain size is comparable 
to the width of the interaction cylinder then the distinction between these 
processes becomes very unclear. 

Most studies of cosmic-ray driven desorption are predicated on the assumption 
that the deposition of energy following a cosmic ray impact leads either to 
`hot spot' heating or, more usually, whole grain heating that dominates 
the desorption process due to thermal evaporation \citep[e.g.][]{HH93} or else
through the triggered release of trapped chemical energy in exothermic reactions
\citep{HABG82,LJO85,SG91,Raw13} leading to total sublimation in ice mantle
`explosions'. 
In addition to promoting efficient ice desorption, cosmic ray-induced reactions 
in the ices between supra-thermal dissociation products may drive the efficient 
formation of complex organic molecules (COMs) \citep{STGH18} and the heating 
process may also enable the efficient diffusion of 
solid-state species, driving a vigorous surface chemistry \citep{K16}.

The astrochemical significance of cosmic ray induced desorption was described 
in the seminal study of \citet{LJO85} in the context of a model in which the 
energy deposition by the passage of a cosmic ray particle, or an X-ray photon, 
through a grain effects thermal desorption of ice mantle species, or else 
triggers the release of trapped chemical energy in an ice mantle `explosion'.
The calculations included the effects of both whole grain heating and spot heating,
the latter being particularly important for the larger grains. To assess the
possible importance of chemical explosions, the study paid particular attention 
to heating events that are capable of yielding $T_{\rm peak}>27$K, this being 
the nominal temperature threshold to enable the mobility of radicals and trigger 
a chemical explosion \citep{HABG82}.

Despite the fact that such cosmic ray impacts are infrequent (typically one
per Myr, when considering `classical' grains of radius 0.1$\mu$m)
the effects on the chemistry were found to be significant.
In their model \citet{LJO85} considered iron nuclei impacting olivine grains, 
although this has been extended in subsequent studies to consider different 
grain compositions and a wider range of cosmic ray types and energies 
\citep[e.g.][]{SGSD04}.

\subsection{The continuous desorption representation}
\label{sec:cont}

The process described above is sporadic, with long intervals between impact events.
This is not particularly easy to incorporate into astrochemical models, 
the subject of which may be environments in which the physical conditions are
evolving on comparable, or shorter, timescales.

In order to provide an easily applicable analytical implementation, appropriate 
for models of time-dependent chemistries, \citet{HH93} -  
hereafter HH93, formulated a continuous representation of the desorption 
process, and this has become the standard model that has been adopted in most 
subsequent astrochemical studies.
As indicated above, the mechanism involves the raising of the grain temperature 
to a level such that the ices can sublimate. The efficiency of the process
thus depends on the binding energies of the various ice species
and the energy and impact rate of the cosmic rays.

As well as the representation of the sporadic process by continuous desorption,
the HH93 representation is based on the following three important simplifications:
(i) the desorption process is entirely dominated by whole grain 
heating/evaporation,
(ii) the grain size distribution for the insterstellar dust particles can be
represented by a single `classical' grain size value of $a=0.1\mu$m, and
(iii) the cosmic ray spectrum can be represented by a single (average) energy 
deposition of $\Delta E\sim$0.4MeV occurring (on average) at intervals of 
$t_{\rm impact}\sim 1$Myr (for a 0.1$\mu$m grain) corresponding to impacts 
by relativistic Fe nuclei with energies in the range 20-70Mev. The energy 
deposition is proportional to the path length through the grain
\citep{LJO85,BJ04}, so that this implies a mean `stopping power' of 20Gev cm$^{-1}$.
With these parameters, HH93 determined that, the temperature of the grain would 
be raised to a peak value of $T_{\rm peak}\sim$70K, assuming WGH. 

The grains then cool by evaporative cooling, caused by 
desorption of ice species (effectively releasing an energy 
greater than or equal to the local binding energy of each ejected molecule).
For the grain size and temperature under consideration this is more efficient 
than radiative cooling and occurs on a very short timescale 
($\sim 10^{-5}$s$^{-1}$), eventually quenching the desorption.
Rather than follow the desorption through any one heating/cooling event, the 
simplification is made that the desorption all occurs at $T=T_{\rm peak}$
for a nominal cooling time ($t_{\rm cool}$); noting that thermal 
desorption is strongly temperature-dependent.
The cycle is then repeated.
For $T_{\rm peak}=70$K, thermal desorption is only significant for the more weakly
bound (volatile) species, such as CO and CH$_4$ and it is important to try and 
quantify the desorption efficiency as accurately as possible when modelling 
the chemistry of these species.

This, of course, is only strictly applicable to ices whose composition is dominated 
by volatile species and is based on a number of important additional
assumptions including:
(iv) the requirement that there are sufficient volatile molecules in the top 
layer of the ice to enable evaporative cooling, 
(v) that the inferred continuous desorption rate for any species does not imply 
a net loss in any one cooling cycle that exceeds the available budget of molecules 
in the ice, and
(vi) that a single binding energy can be used to describe the cooling throughout 
the evaporation process etc. 
The model is therefore predicated on the assumption that one (volatile) species
is the overwhelmingly dominant component of the (surface layers) of
the ice mantles so that it is (a) the principal sublimate, and (b) the dominant 
(evaporative) coolant.
In HH93, this species is taken to be CO, and these assumptions are probably 
satisfied in regions that are very cold, dense and where advanced freeze-out 
has occurred. However the validity of these, and the other assumptions, is 
discussed below.

To represent the episodic process as a continuous desorption rate, 
$k_{\rm crd}^{\rm CO}$ (per bound molecule, per second) the thermal 
evaporation rate (at $T=T_{\rm peak}=70$K) is then multiplied by the
`duty cycle', equal to the cooling timescale ($t_{\rm cool}$) divided by the 
average interval between cosmic ray impacts ($t_{\rm impact}$):

\[ k_{\rm crd}^{\rm CO} = k_{\rm evap}^{\rm CO}.\left[ 
\frac{t_{\rm cool}}{t_{\rm impact}}\right] \]

The rate of thermal sublimation (per molecule) for a zeroth order process 
is given by:
\begin{equation}
k_{\rm evap}^{\rm i} = \nu_0^{\rm i} e^{-E_{\rm B}^{\rm i}/kT_{\rm dust}} 
{\rm ~s}^{-1} 
\label{eqn:kevap}
\end{equation}
where $E_{\rm B}^{\rm i}$ is the binding energy of the adsorbate species $i$,
usually expressed as a temperature ($T_{\rm B}^{\rm i}=E_{\rm B}^{\rm i}/k$), 
$T_{\rm dust}$ is the dust temperature and $\nu_0^{\rm i}$ is the vibration 
frequency of the adsorbed molecule given by 
\begin{equation} 
\nu_{\rm 0}^{\rm i} = \sqrt{2N_{\rm s}kT_{\rm B}^{\rm i}/\pi^2m_{\rm i}} 
\label{eqn:nu0} 
\end{equation}
where $N_{\rm s}$ is the number of binding sites per unit area, and $m_{\rm i}$ 
is the mass of the adsorbed particle. Typically, $\nu_0 = 10^{12}-10^{13}$s$^{-1}$.

Assuming that $T_{\rm dust}=T_{\rm peak}=70$K and $T_{\rm B}=1200$K for CO 
(as specified in HH93), with $N_{\rm s}=10^{15}$cm$^{-2}$ this yields
$k_{\rm evap}^{\rm CO}\sim 3.1\times 10^4$s$^{-1}$, 
and the evaporation timescale for one molecule 
(1/$k_{\rm evap}^{\rm CO}$)$\sim 3\times 10^{-5}$s.
With these values of $T_{\rm dust}$ and $T_{\rm B}$, HH93 estimated that 
the cooling timescale for a grain $t_{\rm cool}=10^{-5}$s$^{-1}$ 
(not to be confused with the evaporation timescale per molecule). So with 
$t_{\rm impact}=3.16\times 10^{13}$s (1 Myr), the duty cycle is 
$\sim 3.16\times 10^{-19}$ and the value for the CO desorption rate as given 
in HH93 is obtained:
$k_{\rm crd}^{\rm CO} = 9.8\times 10^{-15}$s$^{-1}$.

%

Finally, the volumetric evaporation rate (cm$^{-3}$s$^{-1}$) of a species 
$i$ can be determined: 
\begin{equation}
\dot{n}_{\rm i}\, = k_{\rm crd}^{\rm i}\sigma_{\rm H} n_{\rm H} N_{\rm s} f_{\rm i}, 
\label{eqn:crdes}
\end{equation}
where $f_{\rm i}$ is the fractional surface coverage of species $i$,
$\sigma_{\rm H}$ is the grain surface area per hydrogen nucleon,
and $n_{\rm H}$ is the hydrogen nucleon density.
Or, expressed as the rate of change of the fractional abundance of the coolant
species CO ($X_{\rm CO}$):
\[ \dot{X}_{\rm CO}\, = k_{\rm crd}^{\rm CO} \sigma_{\rm H} N_{\rm s},  \]
where we have assumed that $f_{\rm CO}=1$.

Alternatively, the desorption rate can be determined by considering 
the energetics of the process. The advantage of such an approach is 
that (assuming the various assumptions listed above hold) 
the equivalent continuous desorption coefficient is simply defined 
by the number of CO molecules that are required to effect cooling 
($N_{\rm CO}$), and the interval between cosmic ray 
impact/heating events - and is essentially independent of 
the details of the peak grain temperature and cooling 
timescales etc.
Thus, evaporation will continue until a sufficient 
number of CO molecules have been ejected so as to cool the
grain.

Therefore:
\[ N_{\rm CO}.kT_{\rm B}^{\rm CO} \sim f_{\rm E}.\Delta E \]
where $f_{\rm E}$ is the fraction of the total deposited energy 
($\Delta E$) that is removed by evaporative cooling in any 
one cooling cycle.

The cooling rate (whether by evaporative, or radiative cooling) is 
proportional to the grain surface area and the evaporation rate
per molecule \citep[e.g. eqn 3 of][]{LJO85} so that, for a single grain:
\begin{equation}
\Gamma_{\rm CO}= \frac{dE}{dt} = 
k_{\rm evap}^{\rm CO}.4\pi a^2.N_{\rm s}.f_{\rm CO}.kT_{\rm B}^{\rm CO} 
\label{eqn:cocool}
\end{equation}
where $a$ is the grain radius.
The cooling timescale ($t_{\rm cool}$) is then given by:
\[ t_{\rm cool} \sim \frac{f_{\rm E}.\Delta E}{\Gamma_{\rm CO}. } \]
Note that $t_{\rm cool}$ depends on {\em both} the evaporation rate 
and the grain surface area.

The number of molecules that are evaporated from a single grain 
in a period of time equal to $t_{\rm cool}$ is then given by:
\begin{equation}
N_{\rm i} =  k_{\rm evap}^{\rm i}.t_{\rm cool}.4\pi a^2.N_{\rm s} 
=  k_{\rm crd}^{\rm i}.t_{\rm impact}.4\pi a^2.N_{\rm s}. 
\label{eqn:ncool}
\end{equation}

Identifying CO as the coolant species and 
using the values adopted by HH93; $T_{\rm B}^{\rm CO}=1200$K, 
$a=0.1\mu$m, $\Delta E =0.4$MeV and $t_{\rm cool}=10^{-5}$s, 
a value of $f_{\rm E}=0.1$ is implied. 
This gives $N_{\rm cool}^{\rm CO}\sim 3.9\times 10^5$ and corresponds 
to cooling from a peak value of $\sim$70.8K to $\sim$67.1K.
These values of $f_{\rm E}$ and $N_{\rm cool}^{\rm CO}$ have probably 
been underestimated, as discussed in section~\ref{sec:limits} below.

\subsubsection{Desorption rates for other species}
\label{sec:species}

The same formalism described above can be used to determine the equivalent 
desorption rates for other species ($k_{\rm crd}^{\rm i}$) for a given grain
radius ($a$). The cooling and impact timescales  ($t_{\rm cool}$ and 
$t_{\rm impact}$) and hence the duty cycle are fixed, so that - given 
$k_{\rm crd}^{\rm CO}$ - the rates simply scale according to:
\[ \frac{k_{\rm crd}^{\rm i}}{k_{\rm crd}^{\rm CO}} = 
\frac{k_{\rm evap}^{\rm i}}{k_{\rm evap}^{\rm CO}} \] 
and
\[ k_{\rm crd}^{\rm i} = k_{\rm crd}^{\rm CO}
\left[ \frac{T_{\rm B}^{\rm i}}{T_{\rm B}^{\rm CO}} 
\frac{m_{\rm CO}}{m_{\rm i}}\right]^{1/2}
\exp{\left( \frac{T_{\rm B}^{\rm CO}-T_{\rm B}^{\rm i}}{T_{\rm peak}} \right) } \]
The values given in Table~4 of HH93 can be retrieved using this simple 
relationship.

We should, however, note that even assuming the same values for the physical
parameters as in HH93, there is considerable uncertainty in these rates
and some other studies have yielded significantly different values.
So, for example, by considering more localised ice mantle `hot spot' heating 
\citet{BJ04} calculated a rate for H$_2$O desorption that is approximately 30 
times larger than the HH93 value. 
Alternatively, \citet{HC06} applied a detailed continuous-time 
random walk Monte-Carlo approach to the heating of grains and 
the evaporation of molecules to obtain stochastic desorption 
rates.
Using a different (and wider) spectrum of cosmic rays they 
determined larger cosmic ray fluxes (and hence a shorter 
$t_{\rm impact}=10^5$ years), and lower peak temperatures, 
but yielding a significantly larger value of 
$k_{\rm crd}^{\rm CO} = 5.7\times 10^{-13}$s$^{-1}$.  
In addition, there are considerable uncertainties in the rates deriving from 
assumptions about the grain morphology, size distribution, thermal connectivity, 
the nature and spectrum of the cosmic rays etc.

\section{Extension to other grain sizes}
\label{sec:MRN}

\subsection{Dependence on the grain size}
\label{sec:size}

If we accept the `late time, whole grain thermal desorption' physical 
model of HH93 and the various assumptions specified in the previous section
hold, then it is fairly easy to adapt the formalism beyond the `classical' 
scenario. The continuous desorption representation can thus be applied to 
different cosmic ray fluxes and grain sizes, so long as 
25K$\ltappeq T_{\rm peak}\ltappeq$150K. This corresponds to impacts on 
grains with radii in the range $0.03\mu$m$\ltappeq a\ltappeq 0.25\mu$m.

However, the model cannot be extended to very small, or very large grains.
There are several complicating issues for the small grains;
firstly the `granularity' of the stochastic nature of the process becomes more 
important; $t_{\rm impact}$ scales as the inverse of the geometric cross section, so
that for a=0.01$\mu$m, $t_{\rm impact}=100$Myr. This is very much larger than the 
timescale for changes in the physical and chemical environment of the dust grains.
Secondly, for grains smaller than $\sim 0.02\mu$m, the radius of the cosmic ray 
`impact cylinder' is comparable to the grain radius and the desorption efficiency 
may be much higher, due to the increased significance of the direct, impulsive 
desorption mechanisms \citep[e.g. as described by][]{BJ04}.
Thirdly, for the smallest grains, where $T_{\rm peak}\gtappeq $150-160\,K, 
the situation is further complicated by the facts that desorption and cooling by 
the less volatile, but dominant bulk ice component, H$_2$O, may become important,
depending on the assumptions concerning the binding energy and the nature of
the desorption process.
In addition, a greater proportion of the grain will be composed of ice, as opposed
to refractory material, with a different cosmic ray stopping power.
Fourthly, and perhaps most significantly, for the smallest grains
there may be an insufficient number of molecules in the
ice to allow efficient evaporative cooling. Together with the high peak 
temperatures and electron fluxes and the possibility of total grain disruption, 
these factors suggest that for these grains something akin to explosive whole mantle 
desorption may take place.

As small grains dominate the interstellar size distribution the implications are
that the desorption of volatile species may be over-estimated by the
the HH93 formalism (due to the ice budget limitation), whilst that for 
strongly bound species - including complex organic molecules (COMs) - could be 
significantly under-estimated (due to whole mantle desorption/grain disruption).  

For the larger grains (where $T_{\rm peak}\ltappeq 25$K), radiative cooling 
starts to dominate over evaporative cooling, whilst for $a\gtappeq$0.25$\mu$m, 
`cylinder/spot heating' and desorption at the cosmic ray entry/exit 
sites becomes more significant than whole grain heating.
In addition, an additional term in the energy deposition should also be included 
due to the effects of lower energy cosmic rays, whose flux is ill-determined 
in dark cloud cores \citep{LJO85}.  
Both these effects would result in significantly larger desorption rates than 
are predicted by the HH93 model.

However, within the range $0.03\mu$m$\ltappeq a\ltappeq 0.25\mu$m, and
on the continued assumption that CO is the dominant ice component (and 
coolant) in the surface layers, we can easily scale the formulae discussed in 
the previous section to obtain the desorption rate for dust grains of different  
sizes.

As already noted, cosmic rays of a specific energy have an approximately 
uniform `stopping power' or energy deposition rate per unit length (normalised
to a deposition of $\Delta E=0.4$MeV, for a grain with $a=0.1\mu$m in HH93).
Thus
\begin{equation}
\Delta E = 0.4MeV\left( \frac{a}{0.1\mu m}\right) 
\left( \frac{\zeta}{1.3\times 10^{-17}s^{-1}}\right). 
\label{eqn:stopp}
\end{equation}
The temperature-dependence of the specific heat of refractory dust grains is 
reasonably well-known \citep{LJO85} so that $T_{\rm peak}$ for grains of 
different sizes can easily be calculated for an appropriately scaled value of 
$\Delta E$ using the expression;
\[ \Delta E = \frac{4}{3}\pi a^3 \int_{T_0}^{T_{\rm peak}}C_{\rm vol.}T dT, \]
where $a$ is the grain radius, $C_{\rm vol.}$ is the volumic heat capacity
(in J cm$^{-3}$K$^{-1}$, as given by equations 1a \& 1b in \citet{LJO85}),
and $T_0$, $T_{\rm peak}$ are the initial and peak grain temperatures.

We then obtain (for $10<T_{\rm peak}<50$K):
\begin{equation}
T_{\rm peak}^3 = 3.28\times 10^5 \left( \frac{a}{0.1\mu m}\right)^{-2} +T_0^3 
\label{eqn:tmax_lo}
\end{equation}
and (for $50<T_{\rm peak}<150$K):
\begin{equation}
T_{\rm peak}^{2.3} = 1.597\times 10^4 
\left( \frac{a}{0.1\mu m}\right)^{-2} + 4.87\times 10^{-2}T_0^3 + 1990.
\label{eqn:tmax_hi}
\end{equation}

We also note, again assuming that CO is the dominant adsorbate and coolant
species, that for an individual grain, $k_{\rm crd}^{\rm CO}$ and 
$N_{\rm cool}^{\rm CO}$ are simply related through equation~(\ref{eqn:ncool}).
The interval between cosmic ray impacts is inversely proportional to the 
cross-sectional area of the grain, so that
\begin{equation}
t_{\rm impact} = 3.16\times 10^{13} \left( \frac{0.1\mu m}{a}\right)^2
\left( \frac{1.3\times 10^{-17}s^{-1}}{\zeta}\right) 
\label{eqn:timpact}
\end{equation}

Here it is important to note that the energy deposition ($\Delta E$), and hence
$N_{\rm cool}^{\rm CO}$, are proportional to the grain radius ($a$), whilst the interval 
between cosmic  ray impacts is inversely proportional to the grain cross-section 
($t_{\rm impact}\propto a^{-2}$). Hence the CR-induced desorption rate coefficient 
$k_{\rm crd}^{\rm CO}\propto a$.
The same dependency on grain radius can be deduced from 
equation~(\ref{eqn:ncool}) noting that the cooling timescale ($t_{\rm cool}$) for
an individual grain depends on the energy deposition ($\propto a$) and the 
inverse of the grain surface area ($\propto a^{-2}$) as well as the 
evaporation timescale (itself a function of the peak temperature and, 
hence, the grain radius).

This observation is different to what has been deduced/assumed in some recent 
studies \citep[e.g.][]{ZCL18,SZC20} which are based on the erroneous assumption 
that identifies the evaporation timescale for the coolant molecule with the 
cooling timescale for the grain, so that it does not have an explicit grain 
size-depenence.
This was not claimed in HH93 and is self-evidently incorrect (the cooling 
timescale being proportional to the evaporation timescale divided by the 
grain radius, as described above) - e.g. see equation 3 of \citet{LJO85}.
The source of this confusion is almost certainly due to the fact that, for  
0.1$\mu$m grains, and with the parameters used in HH93, {\em both} timescales
are of the order of $10^{-5}$s$^{-1}$.
Obviously, this error is unimportant in the context of models that duplicate 
the parameters used in HH93, but it has serious implications when scaling the
results for different species, grains sizes and cosmic ray spectra etc.
Consequently, compared to previous studies, we obtain somewhat different 
results for the desorption rates averaged over representative grain size 
population distributions.



%
\begin{table}
\caption{Parameters for the models.}
\begin{center}
\begin{tabular}{|l|c|}
\hline
Parameter & Value \\
\hline
Helium abundance ($He/H$) & 0.1 \\
Carbon abundance ($C/H$) & 1.5$\xten{-4}$ \\
Nitrogen abundance ($N/H$) & 7.4$\xten{-5}$ \\
Oxygen abundance ($O/H$) & 2.5$\xten{-4}$ \\
Sulfur abundance ($S/H$) & 1.0$\xten{-7}$ \\
Sodium abundance ($Na/H$) & 1.0$\xten{-8}$ \\
Gas temperature ($T_{\rm g}$) & 10K \\
Initial dust temperature ($T_{\rm d}^0$) & 10K \\
Minimum temperature for evap. cooling ($T_{\rm d}^{\rm min}$) & 25K \\
Cosmic ray ionization rate ($\zeta$) & $1.3\times 10^{-17}$s$^{-1}$ \\
Dust surface area per hydrogen nucleon ($\sigma_{\rm H}$) & 
$6.0\times 10^{-21}$ cm$^{-2}$ \\
Surface density of binding sites ($N_{\rm s}$) & $10^{15}$cm$^{-2}$ \\
Ice layer thickness ($\Delta a$) & 3.7\AA \\
Mean dust grain albedo ($\omega$) & 0.5 \\
\hline
\end{tabular}
\end{center}
\label{tab:params}
\end{table}

\subsection{Application to grain size distributions}
\label{sec:sizedep}

In molecular clouds where the effects of ice grain 
coagulation/agglomeration are not significant, an empirically 
constrained fit to the observed interstellar extinction curve is given
by a simple power law distribution of (uncoated) dust grain sizes
\citep{MRN77}, hereafter MRN, so that the number 
of grains with radii between $a$ and $a+da$ is given by
$n(a)da = n_0 a^{-3.5}$ defined within the approximate limits of
$0.005\mu$m\,$<a<1.0\mu$m.

Ideally, the gas-grain interactions - including the accumulation and 
composition of ices should be followed for a population of dust grains 
of different sizes, weighted according to the grain size distribution.
Unfortunately, the microscopic 
description of freeze-out and desorption of a multi-grain fluid becomes 
extremely complicated once the sporadic nature of the process is 
taken into account, and all the more so in dynamically evolving 
environments where the dynamical timescales may be comparable, 
or less than, the timescale between cosmic-ray heating events 
($t_{\rm impact}$).

However, in the simplified case of assumed continuous desorption from 
homogeneous ice mantles we can obtain order-of-magnitude estimates for the 
population-averaged desorption rates.
For the MRN distribution, and depending on the limits of the distribution,
the value of the implied dust surface area per hydrogen nucleon 
($\sigma_{\rm H}$) lies in the range $3-8\times 10^{-21}$cm$^{2}$.
We adopt a standard value of $6.0\times 10^{-21}$cm$^{2}$ in this study.

We can compare the value of the cosmic ray desorption rate for CO
averaged over this distribution ($k_{\rm crd}^{\rm CO}$), to that obtained 
for a single size grain distribution ($a=0.1\mu$m, as used by HH93) on 
the assumption that both distributions yield 
(i) the same value of the dust-to-gas ratio (by mass), or
(ii) the same surface area per hydrogen nucleon.

Using the assumption of a fixed dust-to-gas mass ratio we note 
that, since the product of $k_{\rm crd}^{\rm CO}.4\pi a^2$
scales as $a^3$, as does the dust mass distribution, then, 
$\overline{k_{\rm crd}^{\rm CO}} = k_{\rm crd}^{\rm CO}(0.1\mu m)$, 
irrespective of the grain size distribution.

Alternatively, assuming that the grain surface area is fixed, 
then we obtain:
\begin{equation}
\overline{k_{\rm crd}^{\rm CO}} = \frac{\int k_{\rm crd}^{\rm CO}(a).4\pi a^2.n(a) da}
{\int 4\pi a^2.n(a) da} 
\label{eqn:k_mrn}
\end{equation}
so that, with $k_{\rm crd}^{\rm CO}(a)\propto a$, this gives:
\[ \overline{k_{\rm crd}^{\rm CO}} = k_{\rm crd}^{\rm CO}(0.1\mu m)
\frac{\sqrt{a_{\rm min}a_{\rm max}}}{a_0}  \]

Using a slightly truncated MRN distribution (for the reasons described 
above), with $a_{\rm min}=0.03\mu$m and $a_{\rm max}=0.25\mu$m
we obtain 
$\overline{k_{\rm crd}^{\rm CO}} = 0.87 k_{\rm crd}^{\rm CO}(0.1\mu m).$
Note, however, that the ratio is close to unity and is fairly 
insensitive to the assumed range of the grain size distribution.
e.g. for $0.02\mu$m$<a<0.25\mu$m:
$\overline{k_{\rm crd}^{\rm CO}} = 0.71 k_{\rm crd}^{\rm CO}(0.1\mu m)$,
whilst for $0.03\mu$m$<a<0.5\mu$m: 
$\overline{k_{\rm crd}^{\rm CO}} = 1.22 k_{\rm crd}^{\rm CO}(0.1\mu m)$.

Previous studies, based on the different understanding of the cooling 
timescale \citep[e.g.][]{ZCL18}, have obtained 
slightly different results for the population-averaged rate for the 
principle desorbate and coolant; CO. Thus 
$\overline{k_{\rm crd}^{\rm CO}} = 1.05  k_{\rm crd}^{\rm CO}(0.1\mu m)$. 
Although, in the case of CO, this discrepancy is small there is a
very significant qualitative difference between the two approaches; 
as pointed out above, studies which do not include an explicit 
grain size-dependence of $t_{\rm cool}$ yield 
$k_{\rm crd}^{\rm CO}\propto a^2$.
Thus, for an MRN-type of grain size distribution, the integrand in 
the numerator of equation~(\ref{eqn:k_mrn}) is $\propto a^{0.5}$ and 
the desorption is therefore weighted towards the larger grains in the 
distribution \citep{SZC20}.
However, we find that $k_{\rm crd}^{\rm CO}\propto a$ and the 
integrand is $\propto a^{-0.5}$ so that it is the {\em smaller} grains
that contribute most to the population-averaged desorption rates. 

This is an important point since, as we have already noted, the physics 
of the desorption process become complex and are ill-defined for the
smallest grains. Additional complications will arise from the fact that,
compared to the larger grains, it is likely that the composition of the 
surface layers of the ices will be quite different for the smallest
grains.
The significance of the smaller grains will be slightly slightly reduced 
due to the fact that for these grains the ice, which has a lower cosmic 
ray `stopping power' than refractory material, makes up a greater 
proportion of the total (ice + refractory) volume of the mantled grains.
This mitigation will not, however, diminish the fundamental conclusion 
that small grains dominate the desorption.
Moreover, it should also be noted that cosmic rays of different energies 
(and with lower stopping powers) will be more able to heat small grains to 
temperatures sufficient for evaporation to occur, further biasing the 
desorption towards the small grain population.

As stated above, the differences are not particularly strong in the case 
of CO, but the situation is obviously much more complicated for other 
(less abundant) species in the ices. 
For these species, the desorption and cooling are not closely coupled 
as they are in the case of CO. In general, they will have different 
binding energies and hence different dependencies of the sublimation 
rate on the peak dust temperature that (unlike CO) are not so strongly 
constrained by the balance between deposition energy and number of 
molecules that are desorbed.
As the smaller grains (with higher peak temperatures) dominate, this 
implies that population-averaged desorption rates for species with 
$E_{\rm B}^{\rm i}>E_{\rm B}^{\rm CO}$ have probably been strongly 
underestimated in previous studies, and we investigate this possibility 
in the next section.

%
%
 
\section{Revision of the continuous desorption model}
\label{sec:limits}

The representation of cosmic-ray induced desorption, which 
is an intermittent, stochastic, process by a continuous desorption 
formulation was implemented for the sake of simplicity and ease 
of use in astrochemical models of time-dependent chemistry where
rate equation descriptions are employed.
We have already noted that several assumptions and simplifications 
have been made, notably that the desorption is primarily driven by
whole grain heating.
The inclusion of other desorption mechanisms (e.g. as described by 
\citet{BJ04}) implies that there is potentially a very wide range in 
the theoretically determined value of $k_{\rm crd}^{\rm CO}$.

However, even within the context of the HH93 model, there are some 
limitations and unqualified assumptions which we identify and discuss 
in this section. 
The issues that we address are:

\begin{enumerate}
\item consideration of limitations to the desorption rates set by the 
availability of the desorbed species and/or coolant molecules, 
\item underestimation of the cooling timescale (and hence the period 
of time during which mantle desorption occurs),
\item the treatment of sporadic desorption as a continuous process,
\item the possibility of species co-desorption, and
\item special consideration of very small grains.
\end{enumerate}
We consider all of these points in this study.
In addition, it should also be noted that the HH93 model was limited 
to consideration of grains of one size, being impacted by cosmic rays 
of a single type and energy. 
This limitation has been addressed in recent studies \citep[e.g.][]{SCI21}
but here we note that, in addition to the additional uncertainties 
concerning the characeristics and attenuation of the cosmic ray energy
spectrum, the dust grain size population, and their thermal properties etc.,
the large uncertainties inherent in the above list may limit the 
value of such studies. 
Therefore, for the purpose of this study, we have not included a 
discussion of these issues.

\subsection{Compositional limitations}

Due to the stochastic nature of the cosmic-ray induced desorption process, 
the desorption rate for any species dictated at the microscopic level and 
is necessarily limited by the energy budget and the availability of that 
species.
Clearly, the continuous desorption model becomes 
meaningless if it effectively translates to a situation in 
which the implied number of desorptions per cosmic-ray impact/event
exceeds the number of molecules in the ice mantle.

As previously noted; for the range of desorption temperatures that we 
consider, evaporative cooling is the dominant cooling mechanism \citep{LJO85}. 
Thus, for a single grain, the product of $k_{\rm evap}^{\rm cool}.t_{\rm cool}$ 
is determined by the requirement that the number of desorbed molecules multiplied by 
their binding energy is some fraction of the energy deposition 
($\Delta E$).
For $\Delta E\sim 0.4$Mev, this equates to $\sim 10^6$ CO molecules in the 
HH93 model (assuming that CO is the dominant volatile component of the ice, and 
therefore the dominant coolant). This is a large number; the formalism therefore 
only holds if there are $\gtappeq 10^6$ CO molecules in the ice before the 
desorption event starts. This is a non-trivial requirement.
%
%
%
%
Continuing with the assumption that CO is the dominant evaporate and coolant,
we can compare $N_{\rm cool}^{\rm CO}$ with the number of available binding 
sites in the surface layers of the ice. Assuming that the total ice thickness
is significantly less than the bare grain radius ($a_0$), which is true for
classical 0.1$\mu$m grains, then:
\[ N_{\rm cool}^{\rm CO} = 4\pi a^2.N_{\rm s}.f_{\rm CO}.n_{\rm L} \]
where $n_{\rm L}$ is the number of ice layers, and $f_{\rm CO}$ is 
the fractional surface coverage of CO.
With the values given in Table~\ref{tab:params} and assuming a pure CO ice 
($f_{\rm CO}=1$) then
\[ N_{\rm cool}^{\rm CO} = 1.26\times 10^6 n_{\rm L} \]

Thus, each desorption event would result in the sublimation of $\sim$1 
monolayer of ice. Indeed if, as discussed below, $t_{\rm cool}$ is significantly
larger than the values used in HH93, this required level of 
desorption may amount to several layers of ice (in the case of classical 
grains) or even exceed the available ice reservoir (in the case of the 
smallest grains).
Throughout this discussion we must remember that the model is only appropriate 
to those regions where there are substantial depositions of ices in which 
CO is the dominant component. This is likely to be true in the darkest, 
coldest and most dense parts of molecular clouds/quiescent star-forming 
regions. 

The situation becomes even more complicated for low abundance volatile 
species which, to satisfy the implied desorption rate, may need to be 
`excavated' from sub-surface levels, raising the issue of the diffusion 
of species from the mantle interiors to the surfaces.
The importance of the `budgetary limitation' can most clearly be seen if we 
consider the desorption of highly volatile species. For example, atomic hydrogen,
which has a very low surface binding energy, will have - according
to the HH93 formulation - a very high value of 
$k_{\rm crd}^{\rm H}\sim 6.0\times 10^{-9}$s$^{-1}$ .
This would effectively mean that, in the continuous desorption representation,
H atoms would be absent from ices. So, in Table~5 of HH93 we see that, by 
balancing the desorption and accretion rates, the ratio of gaseous to surface
abundances of H atoms would be $1.6\times 10^4$.
Since the gas-phase abundance of H atoms in dark clouds is typically 
$\sim 1$cm$^{-3}$, this implies a surface abundance of H-atoms that is so low 
as to preclude the existence of surface hydrogenation reactions.
In reality, so long as the timescale for accretion is less than 
the cosmic ray impact timescale, H atoms would accrete onto cold grains and 
build up a substantial surface abundance for most of the desorption cycle, only
being (efficiently) sublimated at each desorption event. 

We can calculate the {\em maximum possible} value for $k_{\rm crd}^{\rm i}$, 
based on the extreme, yet unlikely, situation of {\em complete} 
mantle desorption.
Here we consider the corresponding equivalent desorption rates 
and then simply state that the continuous desoprtion rates that we have 
calculated above cannot possibly exceed these values.
In that case, the injection of species $i$ (cm$^{-3}$, per desorption
event) is
\[ \Delta n_{\rm i} = (X_{\rm ice}^{\rm i})_0.n_{\rm H}  \]
where $(X_{\rm ice}^{\rm i})_0$ is the fractional abundance of species
$i$ in the solid state, at the time of the desorption event (and, obviously,
assuming that there is no continuous cosmic-ray induced desorption). Thus, 
with the assumption that $n_{\rm H}$ is not time-dependent, this equates to
\[ \dot{n}_{\rm i}\, = \frac{(X_{\rm ice}^{\rm i})_0.n_{\rm H}}{t_{\rm impact}}. \]
Comparing this to the continuous cosmic ray induced desorption rate, given by
equation~(\ref{eqn:crdes}),
%
we obtain
\[ k_{\rm crd}^{\rm i}(max) = \frac{(X_{\rm ice}^{\rm i})_0}{\sigma_{\rm H} 
N_{\rm s} t_{\rm impact}f_{\rm i}} \]
Adopting values from Table~\ref{tab:params} and making the assumptions that
(a) the ice is well-mixed so that $f_{\rm i}$ is approximately equal to 
$(X_{\rm ice}^{\rm i})_0$ divided by the total ice abundance, and (b) most 
of the oxygen has frozen out (predominantly in the forms of H$_2$O or CO),
so that the total ice abundance is $\sim 2\times 10^{-4}$, then we find
\begin{equation} 
k_{\rm crd}^{\rm i}(max) \sim 1.0\times 10^{-12} 
\left( \frac{a}{0.1\mu m} \right)^2 
\left( \frac{\zeta}{1.3\times 10^{-17}s^{-1}}\right) =
\left(\frac{33.3}{t_{\rm impact}}\right)
\label{eqn:kmax}
\end{equation}

It is important to realise that this is an extreme upper limit and so is
quite a stringent constraint. It therefore implies that the values
given in HH93 for the desorption rates for highly volatile species such as 
H, H$_2$, He, C, N, O, CH, NH and NH$_2$ are greatly overestimated.
Indeed, if one adopts the larger value of $t_{\rm cool}$ discussed above and
considers the smaller grains ($a\sim 0.03\mu$m) then this upper limit
will be important for dominant ice components such as CH$_4$, N$_2$, NH$_3$, 
and even CO.
Moreover, we should also be aware that the volatile ice components are probably
more abundant in the outer layers of the ice than they are in the interior,
further reducing the implied value of $k_{\rm crd}^{\rm i}$(max).

\subsection{Extended cooling} 
\label{sec:extend}

In the HH93 study the cooling timescale ($t_{\rm cool}$) was very loosely 
defined and both it and the number of coolant molecules sublimated from
the surface ($N_{\rm cool}$) have probably been significantly underestimated.
The value of $t_{\rm cool}\sim 10^{-5}$s used in HH93 corresponds 
to a drop in temperature of a grain of radius $a=0.1\mu$m from its peak value
of $\sim$70K to $\sim$67.1K, a reduction of the evaporation rate by a 
factor of $\sim 2$, and $\sim$10\% of the energy deposited by a cosmic ray 
impact being lost through evaporative cooling (i.e. $f_{\rm E}\sim 0.1$).
Indeed, by the time that the dust grain temperature has fallen to 60K, some
$\sim 1.3\times 10^6$ molecules will have evaporated and the 
evaporation rate will have fallen to $\sim$6\% of its peak value.

However, even at this level, the implied cooling timescale is still many 
orders of magnitude less than the timescales for either freeze-out, or cosmic 
ray impacts ($t_{\rm impact}$) so that evaporation will continue until radiative
cooling starts to become significant and/or the rate of freeze-out of the 
coolant species becomes comparable to the evaporation rate.
For the classical ($0.1\mu$m) grains that were the basis of the \citet{LJO85}
study radiative cooling only becomes dominant once the grain temperature has 
fallen to $\sim$25 K. More detailed recent studies \citep[e.g.][]{KK20b,SSC21} 
suggest that the threshold temperature may be somewhat higher although, as 
discussed below, the desorption rates are very insensitive to this value.
The rate of freeze-out (cm$^{-3}$s$^{-1}$) of a neutral species, $i$, is given by:
\[ \dot{n}_{\rm i}\, = -n_{\rm H} n_{\rm i}\sigma_{\rm H} S_{\rm i}\left( 
\frac{kT_{\rm g}}{2\pi m_{\rm i}}\right)^{1/2} \]
where $S_{\rm i}$ is the sticking coefficient (typically assumed to be $\sim 1$),
and $T_{\rm g}$ is the gas kinetic temperature.
With the values of $\sigma_{\rm H}$ and $T_{\rm g}$ given in 
Table~\ref{tab:params} the freeze-out timescale for CO 
(when d($n_{CO})\sim n_{CO}$) is thus
\[ \tau_{\rm fo}^{\rm CO}=2.4\times 10^4\left( \frac{10^5 {\rm cm}^{-3}}{n_{\rm H}}
\right) {\rm years} \]
%
%
With our assumption that the coolant species is CO with a binding energy of
$T_{\rm B}=1100$K, the gas temperature is 10K, the sticking coefficient 
$S_{\rm CO}=1$, the ice is composed of CO only and that there has been 
efficient conversion of C to CO in the gas phase, then
the freeze-out and desorption rates are equal when $T_{\rm dust}\sim 20-22$ 
over the gas density range $n_{\rm H}=10^5-10^7$cm$^{-3}$.

We therefore adopt a value of $T_{\rm D}^{\rm min}=25$K in this study. Cooling 
from the peak temperature of $\sim$70K to this value then corresponds to 
$f_{\rm E}\sim 0.95$ - that is to say, evaporative cooling 
can effectively remove nearly all of the deposited impact energy. 
In fact, as shown below, the total time-integrated desorption rate is not at 
all sensitive to the value of $T_{\rm D}^{\rm min}$ as the great majority of the 
evaporation occurs the higher temperatures, due to the exponential form of
the evaporation rate.
In any case, this is significantly larger than the value of $f_{\rm E}\sim 0.1$
adopted in HH93 which implies that the desorption rates may be of the of the 
order of $\sim 10\times$ larger than those determined in that study. 

%
To quantify the total desorption rate, integrated through an impact/cooling cycle,
rather than assume that the desorption of all species occurs at a fixed
temperature ($T_{\rm peak}$) for a single time period ($t_{\rm cool}$), we have
performed calculations that determine the time-dependence of the dust grain 
temperature ($T_{\rm d}$) and hence the number of molecules desorbed as a function 
of time for each species, integrated from $T_{\rm peak}$ to $T_{\rm d}^{\rm min}$. 
This approach is applicable, so long as the cooling timescale is significantly less 
than those for the gas-phase chemistry and freeze-out, or the interval between
cosmic-ray impacts.  
The total number of molecules desorbed from a grain per heating event can then 
be converted to an effective continuous desorption rate using the above formulae. 

The relationships between the grain temperature and energy (in J) can be 
derived from equations (\ref{eqn:stopp}),(\ref{eqn:tmax_lo}) and 
(\ref{eqn:tmax_hi}): For $T_{\rm d}\leq 50$K,
\begin{equation}
T_{\rm d}=1.724\times 10^6 \left(\frac{0.1\mu m}{a}\right) E^{1/3} 
\label{eqn:TE_lo}
\end{equation}
whilst for $50<T_{\rm d}\leq 150$K, 
\begin{equation}
T_{\rm d} = \left[ 2.5\times 10^{17}E\left(\frac{0.1\mu m}{a}\right)^3 
+ 1990.25\right]^{1/2.3}. 
\label{eqn:TE_hi}
\end{equation}

The initial increment in the thermal energy and the peak dust temperature
are defined by the energy deposition per cosmic-ray impact ($\Delta E$).
The initial (pre-impact) temperature of dust grains is taken to be
$T_{\rm d}^{0}=10$K.
For our purposes we assume that the cooling rate (and hence the desorption
rates) are determined on the basis of pure evaporative cooling by a single
coolant species (typically CO) with an invariant binding energy, that dominates
the composition of the surface layers of the ice mantles. As shown below,
this is a reasonably fair approximation.

With this assumption, the evaporation rate (s$^{-1}$) for the coolant species
is given by equation (\ref{eqn:kevap}).
The desorption rate for a single grain is then given by
\[ \dot{n}_{\rm cool}\, \sim k_{\rm evap}^{\rm cool}.4\pi a^2 N_{\rm s} \]
and the cooling rate is given by equation~(\ref{eqn:cocool}) 
(with $f_{\rm CO}=1$).
%
%
Thus, the time-dependence of the dust temperature and the total 
cumulative desorption of the coolant species (and other species, using
the same formalism) can easily be determined.

Table~\ref{tab:cool} shows how the dust grain temperature and number of coolant
molecules that are evaporated ($N_{\rm cool}$) varies with time following the 
impact of a cosmic ray for two values of the coolant molecule binding energy:
(a) 1100K, and (b) 1200K. These calculations have been performed assuming 
`classical' (0.1$\mu$m) dust grains and track the the cooling and desorption of the 
various species from $T_{\rm d} = T_{\rm peak}=70.6$K, down to 
$T_{\rm d} = T_{\rm d}^{\rm min}=25$K. 

\begin{table}
\caption{The time-dependence of the dust temperature and the number
of desorbed coolant molecules per grain, for a grain radius of $a=0.1\mu$m.}
\begin{center}
\begin{tabular}{|lll|lll|}
\hline
 $T_{\rm B}^{\rm cool}=1100$K & & &
 $T_{\rm B}^{\rm cool}=1200$K & & \\
\hline
Time (s) & $T_{\rm dust}$ (K) & $N_{\rm cool}$ &
Time (s) & $T_{\rm dust}$ (K) & $N_{\rm cool}$ \\
\hline
2.3(-7) & 70.6 & 4.2(4) & 9.6(-7) & 70.6 & 3.8(4) \\
1.3(-6) & 69.6 & 2.1(5) & 5.4(-6) & 69.6 & 1.9(5) \\
3.0(-6) & 68.2 & 4.0(5) & 1.3(-5) & 68.2 & 3.7(5) \\
7.9(-6) & 65.7 & 7.7(5) & 3.5(-5) & 65.7 & 7.0(5) \\
5.7(-5) & 58.8 & 1.7(6) & 2.9(-4) & 58.8 & 1.5(6) \\
1.1(-3) & 49.5 & 2.7(6) & 8.1(-3) & 49.5 & 2.4(6) \\
1.9(5)  & 25.0 & 4.0(6) & 1.2(7)  & 25.0 & 3.7(6) \\
\hline
\end{tabular}
\end{center}
\label{tab:cool}
\end{table}


As can be seen from this table, the full cooling timescale (to reach
$T_{\rm d}^{\rm min}$) is quite long ($\sim 10^5-10^7$s) and moderately 
sensitive to the value of $T_{\rm B}^{\rm cool}$ (particularly once 
$T_{\rm d}\ltappeq 50$K), although it is many orders of magnitude smaller 
than the timescale for gas-phase chemical processes, or for the freeze-out of 
molecules onto grains, ($\tau_{\rm fo}^{\rm CO}$) or the interval between 
cosmic-ray heating events ($t_{\rm impact}$).
In any case it is important to note that, in both cases, the great majority 
of the coolant molecules are evaporated within a few milliseconds.
This implies that the assumed value of $T_{\rm d}^{\rm min}$ is not critical
to the determination of the desorption rates.
The table also confirms our observation that the total level of desorption (and
hence cooling) at the nominal value of $t_{\rm cool}=10^{-5}$s adopted by HH93,
is approximately a factor of 10$\times$ less than the value obtained when the 
full cooling to $T_{\rm d}^{\rm min}$ is considered. 

These calculations were repeated for grains with different radii within the 
specified range ($0.03<a<0.25\mu$m), with appropriate scaling for the energy
deposition (as given by equation~\ref{eqn:stopp}), and the interval between 
cosmic ray impacts (as given by equation~\ref{eqn:timpact}).
The total number of of each species that is desorbed 
per impact/heating event ($N_{\rm i}$) can then be converted to an equivalent 
continuous desorption rate using equation~(\ref{eqn:ncool}).
%
%
The values calculated in this way were limited so as to be less than the 
maximum value of $k_{\rm crd}^{\rm i}$(max) given by equation (\ref{eqn:kmax}) above.
Results from these calculations are shown in Tables~\ref{tab:rates_HH93}, 
\ref{tab:rates_1200} and \ref{tab:rates_1100}. 
These give the equivalent desorption rates calculated using the
full cooling calculation for the lower and upper limits of the assumed grain 
size distribution (0.03$\mu$m and 0.25$\mu$m) as well as for the nominal
`classical' grain size of $0.1\mu$m. 

Table~\ref{tab:rates_HH93} shows the results obtained for selected species.
For these calculations, and in most of the models described below, we have used  
the binding energies given in HH93. 
Whilst we recognise that some of these values have been updated
in recent years we have taken this approach for ease of comparison and 
to identify the significance of the effects of extended cooling and/or the 
capping of the desorption rates described above.
However, in the case of H$_2$O, the value used in HH93 ($T_{\rm B}^{\rm H_2O}$=1860K)
is significantly lower (albeit possibly applicable to mixed ices) than the values
currently adopted. We therefore use $T_{\rm B}^{\rm H_2O}$=4800K \citep{Min16}, but
also include results obtained using the HH93 value, designated H$_2$O$^{\dag}$, 
for comparison.
The table includes, again for
comparison, the desorption rates given in Table 4 of HH93.
These were calculated for a grain size of 0.1$\mu$m and so should be
compared to the extended cooling calculations for $a=0.1\mu$m.
Note that some of the values in this table and also tables~\ref{tab:rates_1200} 
and \ref{tab:rates_1100} have been `capped' at the maximum value, as given by
equation~(\ref{eqn:kmax}).

\begin{table*}
\caption{Cosmic ray desorption rates (s$^{-1}$) for selected species
with extended cooling times. $T_{\rm bind}$ is the adopted binding energy 
(in K) for the species (taken from HH93 - but see text in section~\ref{sec:extend} 
for H$_2$O). 
Values are given for grain sizes of 0.01$\mu$m, 0.1$\mu$m, 0.25$\mu$m and 
averaged over an MRN size distribution. Those labelled with a $\star$ are 
upper limits. Rates from HH93 are also given, for comparison.}
\begin{center}
\begin{tabular}{|l|l|lllll|}
\hline
Species & $T_{\rm bind}$(K) & k(HH93) & k(0.03$\mu$m) & k(0.1$\mu$m) & 
k(0.25$\mu$m) & k(MRN) \\
\hline
CO     & 1210 & 9.8(-15) & 2.9(-14) & 9.2(-14) & 1.7(-13) & 7.3(-14)\\
CH$_4$ & 1360 & 1.6(-15) & 1.3(-14) & 8.5(-15) & 2.4(-15) & 1.0(-14)\\
NH$_3$ & 1110 & 5.0(-14) & 9.0(-14)$^\star$ & 9.9(-13) & 5.3(-12) & 9.3(-13)\\
N$_2$  & 1210 & 9.8(-15) & 2.9(-14) & 9.2(-14) & 1.7(-13) & 7.3(-14)\\
SiO    & 3500 & 8.2(-29) & 4.3(-20) & 9.0(-29) & 4.4(-41) & 8.9(-21)\\
H$_2$CO & 1760 & 4.4(-18) & 8.0(-16) & 1.2(-17) & 1.9(-20) & 2.6(-16)\\
CO$_2$ & 2500 & 1.1(-22) & 1.0(-17) & 1.7(-22) & 2.5(-29) & 2.3(-18)\\
HCN    & 1760 & 4.6(-18) & 8.4(-16) & 1.2(-17) & 2.1(-20) & 2.8(-16)\\
HNC    & 1510 & 1.5(-16) & 3.8(-15) & 5.8(-16) & 2.4(-17) & 1.8(-15)\\
HNO    & 1510 & 1.4(-16) & 3.6(-15) & 5.4(-16) & 2.2(-17) & 1.7(-15)\\
H$_2$S & 1800 & 2.4(-18) & 5.9(-16) & 5.9(-18) & 6.0(-21) & 1.9(-16)\\
C$_2$S & 2500 & 9.8(-23) & 8.8(-18) & 1.5(-22) & 2.2(-29) & 2.0(-18)\\
HCS    & 2000 & 1.2(-19) & 1.6(-16) & 2.6(-19) & 2.1(-23) & 4.5(-17)\\
O      & 800 & 3.7(-12) & 9.0(-14)$^\star$ & 1.0(-12)$^\star$ & 6.3(-12)$^\star$ & 9.8(-13)$^\star$\\
OH     & 1260 & 6.3(-15) & 2.5(-14) & 4.7(-14) & 4.7(-14) & 3.8(-14)\\
O$_2$  & 1210 & 9.1(-15) & 2.7(-14) & 8.6(-14) & 1.6(-13) & 6.8(-14)\\
H$_2$O$^{\dag}$ & 1860 & 1.4(-18) & 5.8(-16) & 3.3(-18) & 1.6(-21) & 1.7(-16)\\
H$_2$O & 4800 & - & 6.4(-23) & 1.2(-36) & 4.6(-56) & 1.3(-23)\\
\hline
\end{tabular}
\end{center}
\label{tab:rates_HH93}
\end{table*}

As expected, the extended cooling results in an increase in the rates for
volatile species (such as CO, N$_2$ and NH$_3$) by approximately an order 
of magnitude, whereas the rates for more tightly bound species (such as 
H$_2$O, H$_2$S and CO$_2$) are only larger by a factor of 
$\ltappeq 2\times$. 
For these species, the desorption rates fall very rapidly soon after the 
grain temperature drops from its peak value and little desorption occurs
in the extended cooling period.
For the same reason, the desorption rates for these species are much more
strongly sensitive to the grain size than they are for the volatile species:
the smaller grains have higher peak temperatures so that desorption of 
tightly-bound species can persist for a longer fraction of the cooling 
period.

In these tables we also include the value for the desorption rates
averaged over an MRN grain size distribution, as given by 
equation~(\ref{eqn:k_mrn}).
It should be noted, however, that these values are highly uncertain due to 
(a) the assumed limits of the grain size distribution: we adopt 
$0.03\mu{\rm m}<a<0.25\mu$m, for the reasons given above, but if - for example - 
the lower limit is reduced to $0.01\mu$m, then the mean rates would be reduced 
by a factor of $\sim 2$ for weakly bound species (such as CO and N$_2$) but 
enhanced by an order of magnitude or more for strongly bound species (such as 
H$_2$O and HCS) as one approaches the total desorption limit for the smallest 
grains, and 
(b) the assumption that the composition of the surface layers of the ice is 
independent of the the grain size. In fact, in the case of the simple model
that we investigate below, this approximation is reasonably accurate. 

Looking at the table, we can see that the MRN-weighted desorption rates are 
similar to the rates for 0.1$\mu$m grains for the volatile species but
are very much larger (in some cases by several orders of magnitude) for
the more tightly bound species. As explained above, the rates for these 
species are much larger for the smaller grains in the distribution - and 
the average over the distribution places more weight on these small grains.

As a general result, it should be noted that
the combined effects of including an extended cooling profile and 
averaging over a grain size distribution are to result in significantly
larger values of the desorption rates for both volatile species 
(primarily due to the extended cooling) and tightly bound species 
(due to the contributions from small grains) and to yield a narrower
range of rates than found in previous studies.

Unfortunately, there is is considerable uncertainty in the binding 
energies for the various species, and especially in mixed composition
(predominantly apolar) ices.  
Therefore, in Table~\ref{tab:rates_1200} we give the calculated desorption
rates as a function of binding energy for a species of mass 10\,$amu$, and 
assuming that the mean coolant species binding energy is 1200K.
Whilst the rates depend on both the binding energy ($T_{\rm B}$) and the 
mass ($m_{\rm i}$), the mass-dependence is only present in the vibration 
frequency of the adsorbed molecule 
($\nu_0\propto m_{\rm i}^{-1/2}$; equation \ref{eqn:nu0}).
This may therefore be more useful for astrochemical modelling
purposes; for a specified binding energy, a value of the desorption rate
(either for `classical', 0.1$\mu$m grains, or MRN-averaged) can be taken
from this table and then scaled according to 
\[ k_{\rm crd}^{\rm i} = \left( \frac{10~amu}{m_{\rm i}}\right)^{1/2} 
k_{\rm crd}^{\rm 10amu} \]
(subject to $k_{\rm crd}^{\rm i}\leq k_{\rm crd}^{\rm max}$)
where $m_i$ is the mass of the species (in $amu$).

\begin{table}
\caption{Cosmic ray desorption rates (s$^{-1}$) for a species of mass 10\,amu 
as a function of binding energy for extended cooling times, with 
$T_{\rm B}^{\rm cool}$=1200\,K. Values are given for grain sizes of 0.01$\mu$m, 
0.1$\mu$m, 0.25$\mu$m and averaged over an MRN size distribution.
Those labelled with a $\star$ are upper limits.}
\begin{center}
\begin{tabular}{|l|llll|}
\hline
$T_{\rm bind}$(K) & k(0.03$\mu$m) & k(0.1$\mu$m) & k(0.25$\mu$m) & k(MRN) \\
\hline
$\leq$1100 & 9.0(-14)$^\star$ & 1.0(-12)$^\star$ & 6.3(-12)$^\star$ & 9.8(-13)$^\star$\\
 1200 & 4.9(-14) & 1.6(-13) & 2.9(-13) & 1.2(-13)\\
 1400 & 1.2(-14) & 4.7(-15) & 7.0(-16) & 7.5(-15)\\
 1600 & 3.4(-15) & 2.0(-16) & 2.3(-18) & 1.3(-15)\\
 1800 & 1.0(-15) & 9.5(-18) & 8.5(-21) & 3.3(-16)\\
 2000 & 3.3(-16) & 4.8(-19) & 3.4(-23) & 9.0(-17)\\
 2200 & 1.1(-16) & 2.5(-20) & 1.4(-25) & 2.7(-17)\\
 2400 & 3.5(-17) & 1.3(-21) & 6.0(-28) & 8.2(-18)\\
 2600 & 1.2(-17) & 7.2(-23) & 2.6(-30) & 2.6(-18)\\
 2800 & 3.8(-18) & 4.0(-24) & 1.2(-32) & 8.5(-19)\\
 3000 & 1.3(-18) & 2.2(-25) & 5.2(-35) & 2.8(-19)\\
 3500 & 8.6(-20) & 1.7(-28) & 7.2(-41) & 1.8(-20)\\
\hline
\end{tabular}
\end{center}
\label{tab:rates_1200}
\end{table}

The binding energy of the coolant species (assumed to be CO) is, itself, 
quite uncertain - especially in mixed ices - and so
results obtained for a binding energy of 1100K are shown in
Table~\ref{tab:rates_1100} for comparison. This shows that for species 
with $T_{\rm B}\ltappeq 1200$K, the rates are $\sim 5-10\times$ smaller, 
although for species with larger binding energies the difference is
$\ltappeq 2\times$.

\begin{table}
\caption{Cosmic ray desorption rates (s$^{-1}$) for a species of mass 10\,amu 
as a function of binding energy for extended cooling times, with 
$T_{\rm B}^{\rm cool}$=1100\,K. Values are given for grain sizes of 0.01$\mu$m, 
0.1$\mu$m, 0.25$\mu$m and averaged over an MRN size distribution.
Those labelled with a $\star$ are upper limits.}
\begin{center}
\begin{tabular}{|l|llll|}
\hline
$T_{\rm bind}$(K) & k(0.03$\mu$m) & k(0.1$\mu$m) & k(0.25$\mu$m) & k(MRN) \\
\hline
$\leq$ 1000 & 9.0(-14)$^\star$ & 1.0(-12)$^\star$ & 6.3(-12)$^\star$ & 9.8(-13)$^\star$\\
 1100 & 5.3(-14) & 1.7(-13) & 3.1(-13) & 1.3(-13)\\
 1200 & 2.5(-14) & 2.8(-14) & 1.5(-14) & 2.6(-14)\\
 1400 & 6.9(-15) & 1.1(-15) & 4.3(-17) & 3.3(-15)\\
 1600 & 2.1(-15) & 4.8(-17) & 1.5(-19) & 7.2(-16)\\
\hline
\end{tabular}
\end{center}
\label{tab:rates_1100}
\end{table}

\subsection{Sporadic desorption and ice species co-desorption}
\label{sec:spor}

There is obviously a discrepancy between the sporadic, instantaneous, 
sublimation of the surface layer(s) of ices and its representation by 
a continuous surface 
desorption operating over the entire cycle, during which the physical 
conditions and composition of the surface layers of the ice mantles are 
liable to change.
Notably, if the chemical timescales and in particular, that for freeze-out,
are comparable or longer than the cosmic ray impact timescale then we can
envisage that the effects of successive desorption events may be able
to accumulate. 


Moreover, the notion that the thermal desorption rate of each species
is defined by the single composition binding energy of that species is 
simplistic and unsupported by laboratory data.
These show that the desorption properties of
different species are coupled so that, in many cases, they tend to sublimate 
in several, distinct, `co-desorption bands' \citep{Cetal04,Vetal04,Ob05,Cetal15}.
Indeed, the various binding energies that are reported are highly uncertain 
and are usually defined for pure ice compositions and/or binding to bare 
refractory grain surfaces. 
In the first order evaporative desorption mechanism, the vibration/binding 
characteristics are defined in the context of the molecular environment. 
Thus, the binding energies are not intrinsic but, instead, depend upon the 
composition of, and the degree of segregation within, the ices.
If, for example, we consider a situation 
in which a polar molecule with a large binding energy (such as H$_2$O) resides
as a minority component within a mixed apolar ice, composed of molecules 
with a low binding energy (such as CO), the circumstances that would result 
in the neighbouring CO molecules being sublimated, whilst leaving the 
H$_2$O molecules behind are highly contrived. 
This point was raised in the original study \citet{LJO85}, who suggested that for
ices of mixed volatile/non-volatile composition, desorption would probably occur
through chemical explosions and posssibly fast spot heating.

It would be very difficult to simulate the co-desorption of different species 
in models of continuous desorption as the desorption rates will be critically 
dependent on the ice mantle composition and structure at the time of the 
heating events.

\subsection{Application to very small grains}
\label{sec:vsgs}

The effects of cosmic-ray driven chemistry in cold ices have been shown to be
very significant, especially in the formation of complex organic molecules 
\citep{STGH18}, but a more direct `hot' chemistry may be more applicable in 
the case of very small grains (VSGs), with $a\leq 0.03\mu$m.
As shown in \citet{BJ04}, grains do not have to be very much smaller than the 
classical size of 0.1$\mu$m for prompt and thermal spike sputtering processes
to become significant. Indeed, for the smallest grains, these processes will 
dominate and molecular dynamics models suggest there is a strong possibility 
of refractory core disruption and/or melting \citep[e.g. see Fig 1 of][]{BJ04}.
In addition, the high local electron fluxes and ionizations/dissociations 
will enhance the desorption efficiency and very possibly drive a rich 
gas-phase chemistry, perhaps leading to the formation of COMs.
For these grains, it is therefore apparent that both the peak temperatures 
and also the total desorption efficiencies will be higher than the simple 
scaling laws suggest.

For VSGs the thickness of the ice mantle may be comparable to, or larger,
than the refractory cores, in which case the geometrical increase of the
surface area of the ice layers must be taken into account.
For these grains;
\[ N_{\rm CO} \simeq \frac{4\pi}{3} \left( a^3 -a_0^3\right)
\frac{N_{\rm i}}{\Delta a} \]
where $a_0$ is the (bare) grain radius, $a=a_0+n_{\rm L}\Delta a$
is the the radius of the grain with $n_{\rm L}$ layers of ice and
$\Delta a$ is the thickness of an ice layer.
With the values given in Table~\ref{tab:params}, for VSGs, we find
\[ N_{\rm CO} = 1.13\times 10^5 \left[ (1+0.037n_{\rm L})^3 -1 \right] \]
So, $n_{\rm L}=$20, 30 and 50 correspond to $N_{\rm CO}=4.8\times 10^5,
0.95\times 10^6$ and $2.5\times 10^7$ respectively.
This, of course, underestimates $N_{\rm CO}$ as it does not
take account of the increased path length (and hence energy deposition) for 
cosmic rays passing though grains with relatively thick ice mantles. 
In fact, if we consider the extreme case of $a_0\to 0$, i.e. a pure 
ice grain, then we can identify an `ice grain' radius at which the energy 
deposition is matched by the binding energy of the (assumed pure) ice.
This is:
\[ a = 0.64 \sqrt{f_{\rm ice}/T_{\rm B}^{\rm cool}} \]
where $f_{\rm ice}$ is the energy deposition efficiency (`stopping power')
in ice relative to that for refractory grain material. 
With $f_{\rm ice}\sim 0.5$ and $T_{\rm B}^{\rm cool}=1200$K, this gives 
$a=0.013\mu$m, implying that total mantle desorption will indeed be 
applicable to the smallest grains in a typical MRN-type distribution.
Although this is a very approximate calculation,
even if we include the refractory core in this calculation, then we conclude 
that it is highly likely that for grains with $a\ltappeq 0.02\mu$m
{\em complete} mantle desorption will occur.

Such processes, which involve bulk/volume sublimation, are hard to
quantify by (continuous) surface desorption rates.
Total mantle desorption would be equivalent to a continuous desorption 
injection rate (cm$^{-3}$s$^{-1}$) of:-
\[ \dot{n}_{\rm i}\, = (n_{\rm i})_0/t_{\rm impact} \]
where $(n_{\rm i})_0$ represents the solid-state abundance of species $i$
at the time of the cosmic ray heating event, {\em and in the absence of 
the continuous cosmic-ray heating desorption}. However, $(n_{\rm i})_0$
is not known in the continuous desorption model, so that we need to consider
the chemical processes in individual desorption events to ascertain the 
significance of whole mantle desorption.

%
%

\section{The model of sporadic desorption}
\label{sec:model}

To understand the implications of  the issues that we have raised in the 
previous section; 
(a) a true representation of the sporadic desorption, (b) co-desorption of 
ice mantle species, and (c) total mantle desorption for VSGs, requires a 
different approach to the continuous desorption approximation that is 
normally used. 

To investigate these issues; most 
significantly the differences and discrepancies that may occur as a result of
representing the sporadic heating and desorption processes by continuous desorption
rates we have developed a model that follows the chemical evolution of the 
gas-phase and gas-grain chemistry through several cosmic ray impact cycles.

The model is based on a simple one-point, dynamically static, application of the 
STARCHEM model that we have used in other studies \citep[e.g.][]{RW21}.
For the purpose of our study, we assume that the system is well-evolved and
in chemical quasi-equilibrium, which is consistent with our understanding of
the conditions within dense cores and pre-stellar star-forming regions 
(e.g. Rawlings, Keto \& Caselli, in preparation).
Thus the chemistry is evolved with fixed physical conditions (density, 
temperature, extinction etc.) with the {\sc DLSODE} integration package.
Note that it is considerably more difficult to construct and apply meaningful
models in dynamically evolving systems, particularly those for which the
relevant timescales are short compared to the desorption cycling timescale.

The model is similar to that applied to other dark cloud environments,
the main features of which are as follows:
the time-dependent chemistry is followed of some 81 gas-phase and 17 solid-state 
species, composed of the elements H, He, C, N, O, S, and a representative low 
ionization potential metal; Na, 
coupled through a network of 1250 reactions. Reaction data is taken 
from the UDFA12 database. 
The gas-phase, gas-grain (freeze-out and desorption) and  
surface chemistry 
(limited to the simple hydrogenation of radicals, reactions involving O$_2$ 
and some, empirically constrained, CO$_2$ formation from OH/O/O$^+$ impacting 
bound CO) are all included.
However, for the sake of clarity, and to quantify the role of the 
cosmic-ray induced desorption process in determining the
gas-phase densities in high density/high extinction regions, most
continuous desorption mechanisms (direct and cosmic ray induced 
photodesorption, species-specific enthalpy-of-formation driven desorption, 
non-selective desorption driven by H$_2$ formation etc.) have been 
suppressed in these calculations, with the exception of thermal desorption. 
The model also calculates, on a continuous basis, the detailed, layer-by-layer
composition of the dust grain ice mantles, and also makes due allowance for the
geometrical scaling of the grain size and total dust surface area as ice mantles
accumulate (so that the nominal bare grain radius is an important 
parameter); the so-called three-phase model of chemical evolution. Thus, 
desorption is assumed to only be effective from the surface layer of ice, which
includes the partially complete top layer, and any exposed parts of the
lower layers. 
Only the surface layers are treated 
as being chemically active, and liable to addition or removal.
In this simplified model we do not include allowances for grain porosity or
the diffusion between layers or account for the possibility of ice melting 
and the mixing of the layers that may occur as a result of strong heating 
events.

The (undepleted) elemental abundances are given in Table~\ref{tab:params} 
and are typical of values adopted for low mass star-forming regions
\citep[e.g.][]{vD21} although, as we are not attempting to model a specific source
in this study, the values are not critically important.  
The chemical initial conditions are taken to be atomic, with the H$:$H$_2$ ratio 
set to 10$^{-3}$ and we assume that $\gtappeq$90\% of the carbon has been 
converted to CO at the start of the calculations. Other parameters are as 
specified in Table~\ref{tab:params}

There are three phases in the models:
\begin{itemize}
\item[(I)] Initial conditions are established by evolving the chemistry for a 
long period of time ($\sim$ 10\,Myrs).
\item[(II)] The desorption phase, during which there is a rapid return of a 
fraction of the ice mantle to the gas-phase, following one of the desorption
models described below. The timescales for these processes are very much smaller than
the timescales of Phases I or III and so the return to the gas-phase is treated as 
being instantaneous.
\item[(III)] The chemistry is followed (again, with static physical conditions)
for a period of $t=t_{\rm impact}$, as given by equation~\ref{eqn:timpact} after 
which Phase II is repeated.
\end{itemize}
Phases II and III are repeated for 5 cycles, or until there are no cycle-to-cycle
variations in the abundances.

To investigate the various issues discussed above we have considered several different
representations of the cosmic ray desorption process that operate in Phase II, 
in three variants of our model:

\subsection{Model 1: continuous desorption}
\label{sec:model1}

As a benchmark, and for the sake of comparison, we have
included models in which the (surface) cosmic-ray induced desorption is 
treated as being continuous and operating throughout all stages of chemical 
evolution (so that Phase II is effectively absent). 
This is essentially the same representation as described in HH93 {\em et seq.} 
so that for these models we apply the formalism and adopt the parameters used in 
HH93 scaled, where necessary, for different grain sizes/size distributions. 

We have considered three variations of this model:
for Model 1a, we use the same binding energies and desorption rates 
as given in HH93, and do not impose the upper limits to the rates as described 
in the previous section. This is therefore a nearly exact simulation of the HH93 model.
For this reference model, we adopt all other parameters as used by HH93 for
classical ($a=0.1\mu$m) grains and use desorption rates as determined by the
formulae above. 



In general, we can scale the cooling timescale, as used in HH93, 
for an arbitrary grain size:
\[
t_{\rm cool} = 10^{-5}\left( \frac{0.1\mu m}{a}\right) 
\exp{ \left[ T_{\rm b}^{\rm cool}\left( \frac{1}{T_{\rm peak}}-\frac{1}{70.84}\right) \right]}
\]
where $T_{\rm B}^{\rm cool}$ is the binding energy (in K) of the coolant
species.
Again, note that this is a function of the grain size, both explicitly in the 
$(1/a)$-dependence and implicitly in the size-dependence of the peak 
dust grain temperature and hence the (single molecule) evaporation timescale.

For Models 1b and 1c we use the  desorption rates calculated with extended 
cooling, as discussed and calculated in the previous section, and we also set the
upper limits to the desorption rates using equation~(\ref{eqn:kmax}).
In Model 1b, we adopt a single nominal grain size of 0.1$\mu$m, 
whilst in Model 1c we use the desorption rates averaged over an MRN grain 
size distribution. The rates for the key species are given in  
Table~\ref{tab:rates_HH93} for both models.


For the other models described below, the continuous cosmic-ray induced 
desorption is replaced by one of the sporadic desorption mechanisms in 
Phase II, with no desorption (other than thermal desorption) operating in 
either of Phases I or III.

\subsection{Model 2: sporadic, species-specific desorption}
\label{sec:model2}

In this model we assume sporadic, stochastic desorption. 
That is to say, there is (an assumed instant) injection of 
each species immediately after a cosmic-ray impact/heating event.
This is therefore closer to the `true' situation than the continuous
desorption simplification employed in HH93 and Model 1.
For clarity, we have again assumed a single `classical' (0.1$\mu$m) 
grain size. 

Following other studies, in this model we assume that the thermal desorption 
characteristics of each species are defined for pure ices and are independent of
the chemical/physical environment of the desorbing molecules.  Therefore,
for the purpose of this model (only) we have assumed that - as in the 
HH93 model - the composition of the ice is homogeneous, i.e. there are no 
compositional variations in the various ice layers.   
This is obviously a major simplification but,
%
%
if we were to adopt a microscopic layer-by-layer approach to the ice 
composition (as in our other models), then the inconsistencies due to differential
desorption as discussed in section~\ref{sec:spor} would be evident.
For the order-of-magnitude type of calculations that we are considering in this
model the assumption of a homogeneous ice composition is therefore more 
workable and is acceptable for our purposes. 

The desorption rates derived from the extended cooling
calculations are used to determine the number of atoms/molecules of each 
species that are released into the gas-phase. Thus, the incremental change to 
the gas-phase fractional abundance of species $i$ is given by:
\[ \Delta X_{\rm i} = k_{\rm evap}^{\rm i}.t_{\rm cool}.\sigma_{\rm H} N_{\rm s} 
F_{\rm i} = k_{\rm crd}^{\rm i}.t_{\rm impact}.\sigma_{\rm H} N_{\rm s} F_{\rm i} \]
where $k_{\rm crd}^{\rm i}$ are the equivalent cosmic-ray induced desorption
rates, calculated as described in section~\ref{sec:extend} above.
$F_{\rm i}$ is the (bulk) fraction of the ice mantle that is composed of species $i$
which, in this model, is identical to the fractional surface coverage of the 
species, $f_{\rm i}$.
The incremental desorption, so calculated, is subject to the obvious constraint 
that the number of molecules desorbed per heating event cannot exceed the total
initial number of molecules in the ice (i.e. $\Delta X_{\rm i} \leq X_{\rm i(ice)}$). 

\subsection{Model 3: Sporadic co-desorption}
\label{sec:model3}

The other extreme to the species-specific desorption adopted in Model 2
is to assume that the desorption process does not discriminate between different 
species, so that a single common binding energy applies to {\em all} 
surface species.
This corresponds to the case of complete co-desorption of ice species.
This is, of course, highly speculative and will depend on a number of unknown
factors, such as the degree of compositional segregation within ice layers but is, 
to some degree, guided by laboratory evidence.
To simulate this in the model we assume that complete layers, 
or a fraction of a layer, of ice are completely sublimated without consideration
of the compositional variations within those layers.
To determine the number of layers that are sublimated, we can use the same 
formalism as employed in the previous section; simply equating the total 
sublimation energy with the energy that is required to cool the grain.
This model is, again, a huge over-simplification but, together with Model 2, 
it serves to demonstrate the range of desorption efficiencies that may be possible.

For the calculations presented here we assume that 
$T_{\rm B}^{\rm i} = T_{\rm B}^{\rm cool}=1210$K,
equal to the binding energy for CO in the HH93 model.
Whilst this `common binding energy' is ill-defined it should be noted that, 
with this formalism, uncertainties in its value will only result in variations in the
desorption efficiencies of order unity.

Obviously, for this model, the actual desorption that occurs in each cycle 
depend on the composition of the surface layers and so the compositional 
structure of the ices is very important. We consider two scenarios:
\begin{itemize}
\item[(a)] Desorption from `classical' ($0.1\mu$m) grains, for which a 
limited number of ice layers will be sublimated as described above, and
\item[(b)] Desorption from VSGs ($0.01\mu$m), for which {\rm complete} mantle 
sublimation, of all ice layers, occurs - as discussed in section 
\ref{sec:vsgs}.
\end{itemize}
For Model 3(a), the derived desorption rate for species with a binding energy
of $\sim 1100-1200$K from `classical' grains with $a=0.1\mu$m
corresponds to the complete desorption of the top $\sim$3 layers of the ice, 
which is what we have adopted for the model. 

\subsection{Model parameters}
\label{sec:parameters}

For each of our models,
calculations have been performed for two representative gas densities and 
extinctions; (i) $n=10^5$cm$^{-3}$, $A_{\rm v}=10$, and 
(ii) $n=10^7$cm$^{-3}$, $A_{\rm v}=50$; which are typical of values within prestellar
dark cores.
As we have justified above,
for most models the calculations have been performed for `classical' ($a=0.1\mu$m)
grains, for ease of comparison with previous studies. Model 1b is for an averaged
MRN grain size distribution and Model 3b is for VSGs ($a=0.01\mu$m).
Where $a=0.1\mu$m, the interval between cosmic ray impacts is $t_{\rm impact}=1.0$\,Myr, 
whereas for $a=0.01\mu$m, $t_{\rm impact}=100.0$\,Myr. The energy deposition per
cosmic ray impact is taken to be proportional to the grain size, $a$.

Of course, these models determine the time-dependences of the chemical 
abundances (gas-phase and solid-state) for a single representative grain 
through one impact/cooling cycle. 
To ascertain the significance of cosmic-ray induced desorption in astrochemical
models we obviously need  to obtain the mean abundances for a system containing 
many grains at different phases of the cycle. We therfore determine statistical averages 
of both the gas-phase and solid-state abundances by calculating the (time-averaged) 
mean values over individual cycles.
The key characteristics of the models are given in Table~\ref{tab:models} and
the values of other parameters are as specified in Table~\ref{tab:params}.

\begin{table}
\caption{Description of the models}
\begin{center}
\begin{tabular}{|l|l|}
\hline
Model & Desorption \\
\hline
Model 1a & Continuous, HH93 rates \\
Model 1b & Continuous, extended cooling, $a=0.1\mu$m  \\
Model 1c & Continuous, extended cooling, MRN-averaged \\
Model 2 & Sporadic, evaporative cooling \\
Model 3a & Sporadic, co-desorption of top 3 ice layers \\
Model 3b & Sporadic, total ice desorption for VSGs ($a=0.01\mu$m) \\
\hline
\end{tabular}
\end{center}
\label{tab:models}
\end{table}

\section{Model results}
\label{sec:results}

In this section we present the results from the models as plots of the 
abundances of key species as a function of time for each of the five studied 
desorption cycles, and as the numerical time-averaged values of the 
abundances for the final ($\sim$equilibrium) cycle.
We have considered the chemical evolution following the impact(s) of
cosmic rays on a single, representative, grain. However, the  
time-averaged values of the abundances in the limit cycle are effectively 
representative of the mean values of the abundances in an ensemble of grains
at different stages within the heating and desorption cycle.
This is, of course, subject to the limitation that we are considering a single, 
or in some cases population-averaged, value of the grain size and a single cosmic
ray energy, but does serve to show the effects in macroscopic objects.
We must also make it clear that
these results have been obtained for a 
rather specific physical model with somewhat contrived chemical constraints 
and are not generic. As such, and unlike the the more rigorous discussion
in section~\ref{sec:limits}, they must be considered indicative, rather than 
definitive.

\begin{figure}
\includegraphics[width=1.0\columnwidth,trim=0.7in 5.7in 4.1in 1in]{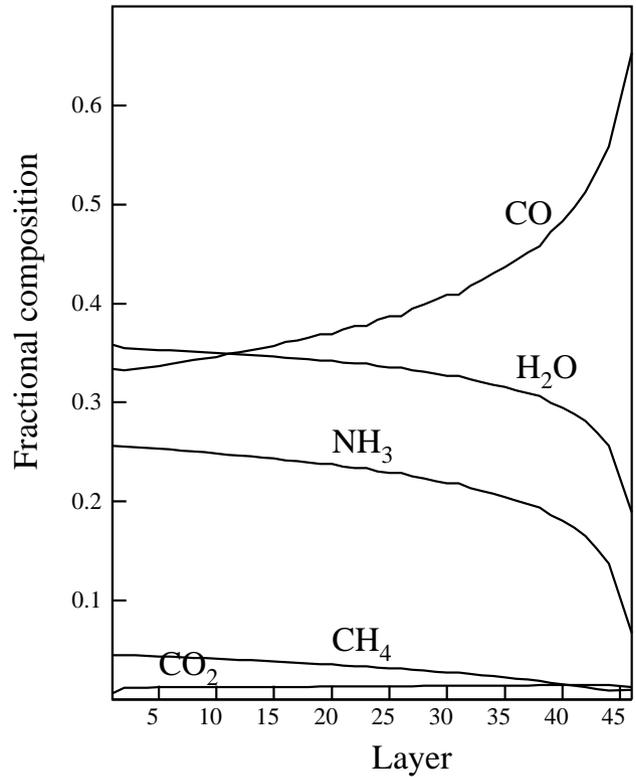}
\caption{Ice mantle composition as a function of monolayer number, for models with
a=0.1$\mu$m and n=10$^5$cm$^{-3}$.}
\label{fig:results_fig1}
\end{figure}

Firstly, in Figure~\ref{fig:results_fig1} we show the composition of the ice 
mantles as determined at the end of Phase I and for the case of grains of 
radius $a=0.1\mu$m and a gas density of $n=10^5$cm$^{3}$.
Abundances are shown as a function of monolayer, so that the values on the 
right hand side of this plot give the composition of the outermost (surface) 
layer. 

From this figure we can see that, for the conditions that we have 
investigated in this study, the composition of the outer 5-10 layers 
of the ice mantles is dominated by CO. 
The surface layers are composed of CO ($\sim$70\%) and
H$_2$O ($\sim$20\%) with other species (e.g. NH$_3$, CH$_4$ and CO$_2$) 
accounting for the remaining 10\%.
Deeper into the ice (i.e. for the inner 25 layers), the composition is
much more heterogeneous.
In our model we have adopted somewhat arbitrary physical and chemical 
conditions: we follow the chemical evolution and ice formation whilst 
keeping the physical parameters (such as the density and temperature) static, 
and exclude all desorption mechanisms, other than thermal desorption 
and cosmic-ray induced desorption.
Making these assumptions, we find that the ice mantle 
composition as shown in this figure is fairly insensitive to the assumed 
mean grain size and gas density
(over the ranges $0.01\mu m<a<0.1\mu m$, $10^5$cm$^{-3}<n<10^7$cm$^{-3}$).
This will almost certainly not be the case for more realistic 
chemical/dynamical simulations. However, for our models, this justifies
{\em a posteriori} the simplificiation made above that we
can adopt a single species (CO) and binding energy for the grain cooling 
calculations.

\begin{figure*}
\includegraphics[width=0.93\columnwidth,trim=0.7in 5.7in 4.1in 1in]{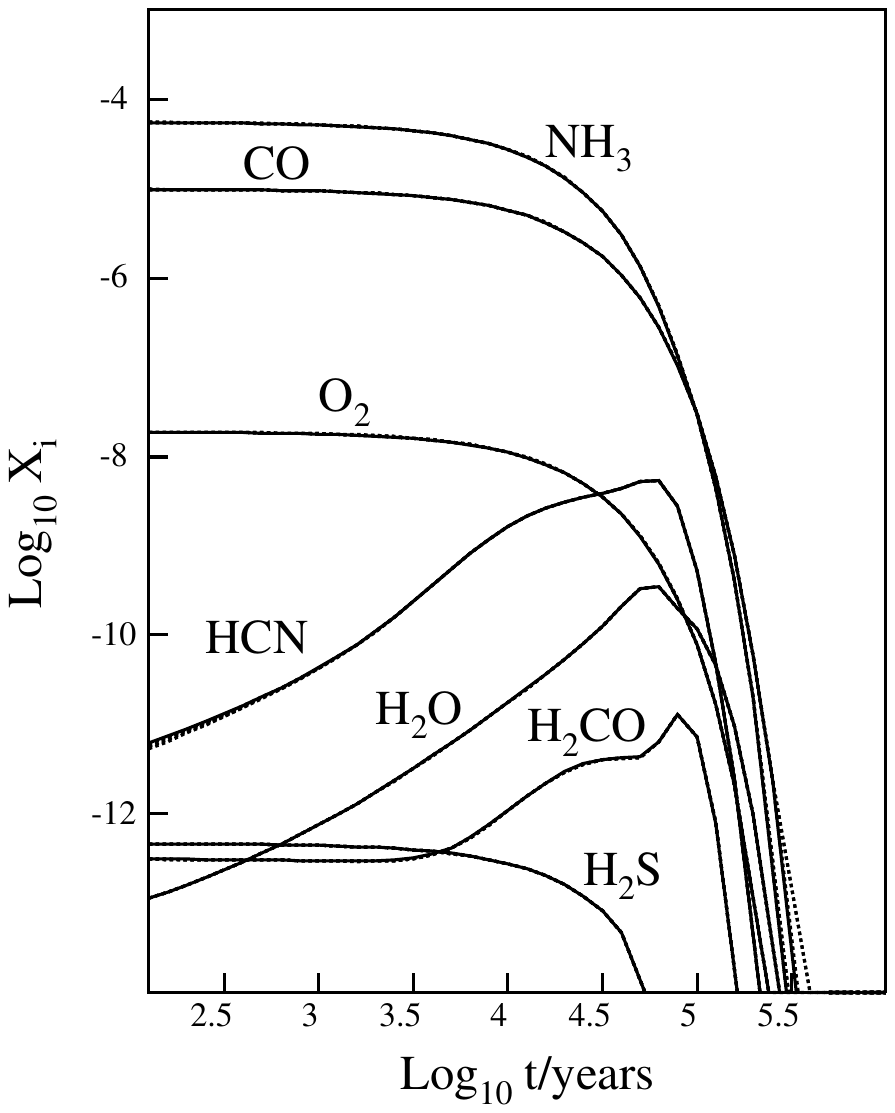}\hfill
\includegraphics[width=1.0\columnwidth,trim=0.7in 5.7in 4.1in 1in]{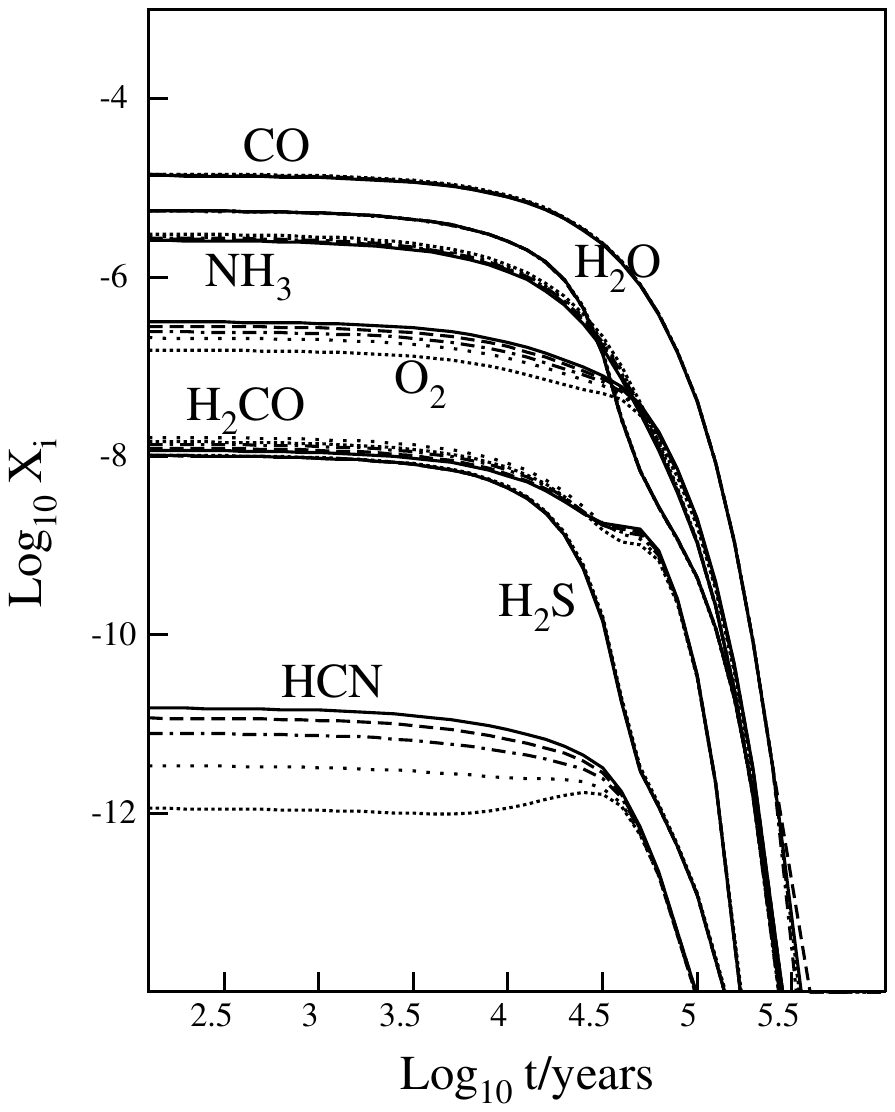}
\caption{Results from models 2 (lhs) and 3a (rhs), with 
a=0.1$\mu$m, n=10$^5$cm$^{-3}$: Fractional abundances of selected species are
shown as a function time since the cosmic ray heating event for the first 
five impact cycles. For model 3a, the logarithmic abundances of HCN have 
been lowered by 4 for the sake of clarity. 
The solid lines show the results for the final (fifth) 
cycle, whilst the results from the previous cycles are shown by the close dotted 
(cycle 1), wide dotted (cycle 2), dash-dotted (cycle 3) and dashed (cycle 4) 
lines.}
\label{fig:results_fig2}
\end{figure*}

Figure~\ref{fig:results_fig2} shows the time-dependence of several key 
gas-phase species for each of the five cycles in the cases of 
Model 2 (sporadic desorption; Fig~\ref{fig:results_fig2}a) and Model 3a 
(sporadic co-desorption; Fig~\ref{fig:results_fig2}b).
%
From this figure we can see that the modelled gas-phase chemical abundances are 
clearly different for the two models so
that, for example, NH$_3$ is very efficiently produced in Model 2 (sporadic
desorption) whereas the co-desorption in Model 3a results in strong enhancements 
of the more tightly bound species, such as H$_2$O and H$_2$S.
We can also see that the chemical behaviours can be quite complex 
and that we need to consider the time-averaged abundances in each cycle to obtain 
a meaningful understanding of the chemical trends.
The behaviour of H$_2$O (and HCN) in Fig~~\ref{fig:results_fig2}a is 
particulary interesting; the desorption of these species is minimal, yet they
experience strong second order transient enhancements that are driven by the 
enriched gas-phase chemistry.

It is also apparent that, for the particular values of the free parameters used in 
these models, there are very little cycle-to-cycle variations in the abundance 
profiles (except for Model 3a in the case of HCN and, to a lesser extent, NH$_3$).
This is primarily a consequence of the fact that the gas-phase and freeze-out 
timescales are relatively short compared to the interval between cosmic
ray impacts ($t_{\rm impact}$), so that the transient chemical effects
of the desorption in one cycle are `lost' before the next cycle 
starts.  

Table~\ref{tab:results_loden} presents the time-averaged gas-phase abundances
of selected species for each of the models described above, 
for the final (fifth) cycle of the calculations. Results are shown for each of 
the desorption models (Models 1-3) for an environment in which the density
is $10^5$ cm$^{-3}$ and the visual extinction is 10 magnitudes.
Table~\ref{tab:results_hiden} shows the results (with the same format as 
Table~\ref{tab:results_loden}) for a higher density ($10^7$ cm$^{-3}$) and
darker ($A_{\rm v}=50$) environment.

Comparing the results for the continuous desorption model with extended cooling 
(Model 1b) to the reference model (Model 1a) we can see that, for the low density 
case, the differences are mostly
fairly minimal (although non-negligible in the cases of CH$_4$, H$_2$CO and HNO)
indicating that continuous CR-induced desorption is not critical in defining the 
chemistry in these environments.
In contrast, for the higher density case, the abundances are typically enhanced by a 
factor of $\sim$10 or more - as expected. 
Extending the model from a single size ($0.1\mu$m) population to an average 
over an MRN-type size distribution (Model 1c) results in strong enhancements of the 
more tightly-bound species (such as H$_2$CO, and the sulfur species) and
these are particularly significant in the high density case. 
Such effects are a consequence of the strong contributions from the smaller grains,
that are heated to higher temperatures by cosmic ray impacts.
The differences between the sporadic desorption model (Model 2) and the continuous
desorption model (with extended cooling; Model 1b) are mostly fairly minor - due 
to the relatively short chemical timescales for the adopted model parameters, 
as discussed above. 
However, there are a few notable differences for the low density 
calculations (only); importantly, Model 2 results in large enhancements of CO, 
NH$_3$ and HNO (by factors of $\sim$10, $\sim$41, and $\sim$24 respectively).   

The (non species-specific) co-desorption model (Model 3a) results in very strong
enhancements, by factors of $\sim 10-100$, for nearly all species (other than CH$_4$ 
and NH$_3$). This is particularly acute in the cases of 
those that have small, yet significant, abundances in the ices - such as H$_2$O, 
CO$_2$ and the sulfur species - which are enhanced by several 
orders of magnitude. As with the other models, the trends tend to be amplified 
in the high density calculations, although not for CO, CH$_4$, NH$_3$, N$_2$ 
or O$_2$.

The model specific to total mantle desorption from VSGs, with $a=0.01\mu$m
(Model 3) again results in big changes to the abundances but, whilst some 
(such as H$_2$O, CO$_2$ and the sulfur species) are very strongly enhanced - by 
several orders of magnitude in the high density calculations - the 
more volatile species (CH$_4$, NH$_3$, N$_2$, O$_2$, HCN, HNC and even CO in the 
high density case) actually have lower abundances than those determined in the 
extended cooling continuous desorption model (Model 1b).
This indicates that a degree of chemical differentiation may occur in those 
situations where a large population of small grains dominates the desorption.

%
%
%
%

\begin{table*}
\caption{Model results (low density; n=10$^5$cm$^{-3}$, 
$A_{\rm v}$=10): 
time-averaged fractional abundances of selected species for the final cycle.
For H$_2$O, results are given for two values of the binding energy 
(see section~\ref{sec:extend} for details).}
\begin{center}
\begin{tabular}{|l|llllll|}
\hline
Species & Model 1a & Model 1b & Model 1c & Model 2 & Model 3a & Model 3b \\
\hline
CO     & 1.8(-8) & 2.0(-8) & 2.5(-8) & 2.0(-7) & 2.7(-7) & 1.6(-8) \\
H$_2$O & 3.6(-11) & 3.8(-11) & 4.7(-11) & 2.3(-11) & 6.9(-8) & 1.1(-8) \\
H$_2$O$^{\dag}$ & 3.6(-11) & 3.8(-11) & 8.0(-11) & 2.8(-11) & 6.9(-8) & 1.1(-8) \\
CH$_4$ & 4.7(-10) & 4.5(-9) & 5.1(-9) & 1.1(-9) & 1.2(-8) & 1.1(-9) \\
NH$_3$ & 1.1(-8) & 2.1(-8) & 2.7(-8) & 8.6(-7) & 3.5(-8) & 7.3(-9) \\
N$_2$  & 4.7(-10) & 5.7(-10) & 7.9(-10) & 2.5(-9) & 6.4(-9) & 8.5(-11) \\
O$_2$  & 2.9(-10) & 1.4(-10) & 2.3(-10) & 4.0(-10) & 7.6(-9) & 1.4(-11) \\
H$_2$CO & 1.2(-12) & 6.1(-12) & 9.2(-12) & 7.8(-13) & 2.4(-10) & 5.3(-12) \\
CO$_2$ & 7.7(-14) & 9.4(-14) & 2.9(-13) & 2.1(-13) & 9.8(-9) & 6.3(-10) \\
HCN    & 2.0(-11) & 5.7(-11) & 1.2(-10) & 3.7(-10) & 3.1(-9) & 2.4(-11) \\
HNC    & 2.3(-11) & 6.4(-11) & 1.2(-10) & 4.8(-10) & 2.3(-9) & 2.7(-11) \\
HNO    & 1.7(-12) & 9.4(-12) & 2.3(-11) & 2.3(-10) & 2.0(-9) & 2.0(-11) \\
H$_2$S & 4.3(-16) & 8.4(-16) & 2.9(-14) & 9.1(-15) & 1.2(-10) & 1.3(-11) \\
C$_2$S & 2.1(-25) & 4.8(-25) & 6.4(-21) & 5.7(-24) & 1.2(-16) & 3.6(-16) \\
HCS    & 2.7(-19) & 9.6(-19) & 7.5(-17) & 1.9(-18) & 2.0(-12) & 1.0(-13) \\
\hline
\end{tabular}
\end{center}
\label{tab:results_loden}
\end{table*}

\begin{table*}
\caption{Model results (high density; $n=10^7$cm$^{-3}$, $A_{\rm v}$=50): 
time-averaged fractional abundances of selected species for the final cycle.
For H$_2$O, results are given for two values of the binding energy 
(see section~\ref{sec:extend} for details).}
\begin{center}
\begin{tabular}{|l|llllll|}
\hline
Species & Model 1a & Model 1b & Model 1c & Model 2 & Model 3a & Model 3b \\
\hline
CO     & 3.1(-10) & 2.7(-9) & 2.2(-9) & 2.0(-9) & 2.9(-9) & 1.6(-10) \\
H$_2$O & 9.1(-16) & 8.6(-15) & 6.9(-15) & 2.5(-16) & 9.3(-10) & 1.1(-10) \\
H$_2$O$^{\dag}$ & 8.2(-15) & 2.5(-14) & 8.2(-13) & 4.6(-14) & 9.3(-10) & 1.1(-10) \\
CH$_4$ & 4.7(-13) & 6.9(-12) & 6.8(-12) & 1.1(-11) & 3.6(-11) & 1.1(-11) \\
NH$_3$ & 2.0(-10) & 3.7(-9) & 3.5(-9) & 9.0(-9) & 5.3(-10) & 7.3(-11) \\
N$_2$  & 1.8(-12) & 1.3(-11) & 1.1(-11) & 8.1(-12) & 1.7(-11) & 4.3(-13) \\
O$_2$  & 9.5(-14) & 8.1(-13) & 6.4(-13) & 6.0(-14) & 5.8(-13) & 2.3(-15) \\
H$_2$CO & 2.0(-17) & 8.8(-17) & 1.3(-15) & 1.1(-17) & 2.4(-13) & 2.6(-15) \\
CO$_2$ & 1.8(-19) & 3.8(-18) & 1.8(-15) & 1.3(-18) & 8.0(-11) & 6.1(-12) \\
HCN    & 6.3(-16) & 1.6(-14) & 2.7(-14) & 9.8(-15) & 4.3(-12) & 5.7(-14) \\
HNC    & 2.3(-15) & 5.0(-14) & 5.9(-14) & 6.5(-14) & 2.8(-12) & 2.9(-14) \\
HNO    & 2.9(-16) & 2.8(-15) & 5.5(-15) & 1.3(-15) & 1.2(-13) & 4.6(-16) \\
H$_2$S & 5.2(-17) & 1.2(-16) & 3.7(-15) & 1.0(-16) & 2.5(-12) & 1.3(-13) \\
C$_2$S & 1.6(-25) & 1.8(-25) & 2.5(-21) & 1.0(-25) & 9.2(-17) & 6.0(-18) \\
HCS    & 7.4(-20) & 1.3(-19) & 2.2(-17) & 3.6(-20) & 4.8(-14) & 6.2(-16) \\
\hline
\end{tabular}
\end{center}
\label{tab:results_hiden}
\end{table*}

\begin{figure}
\includegraphics[width=0.93\columnwidth,trim=0.7in 5.7in 4.1in 1in]{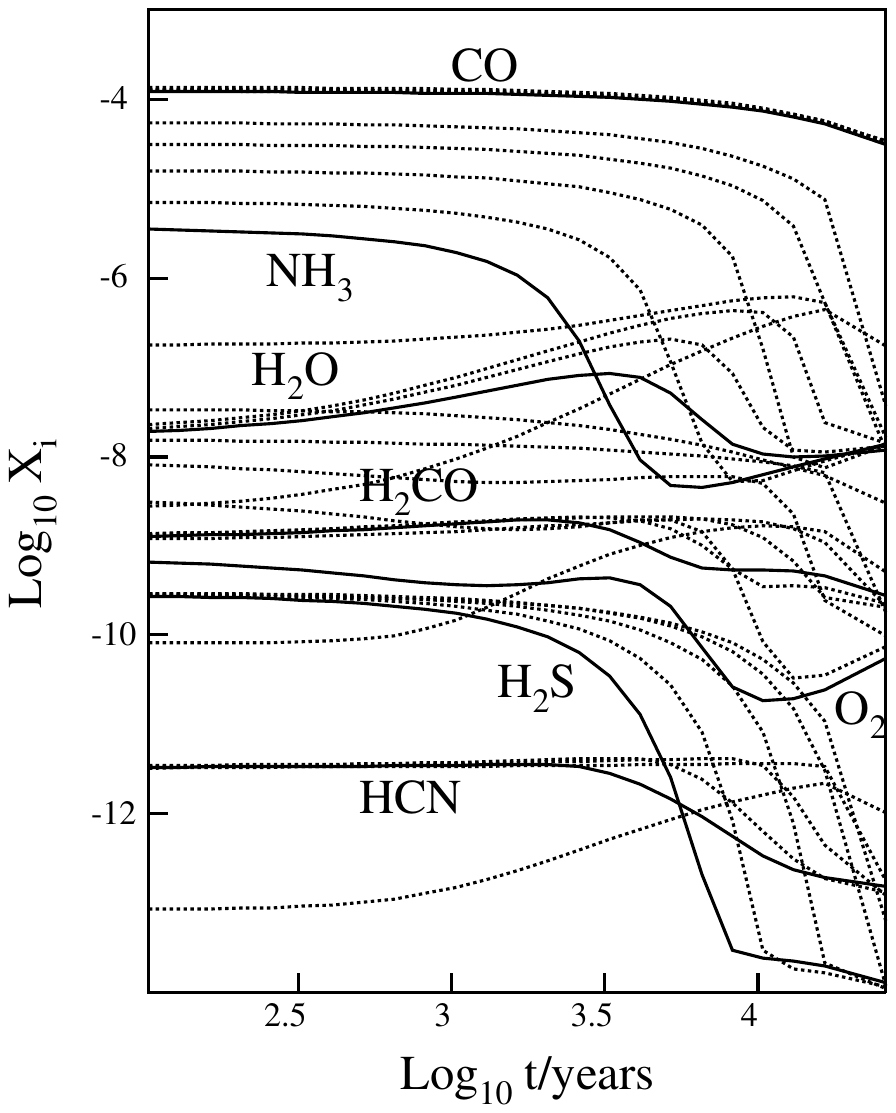}
\caption{Results from Model 2, with an enhanced value of $\zeta$
($5\times 10^{-16}$s$^{-1}$), for $a=0.1\mu$m, $n=10^5$cm$^{-3}$ and 
$A_{\rm v}=10$: Fractional abundances of selected species are
shown as a function time since the cosmic ray heating event for the first 
five impact cycles.
The logarithmic abundances of O$_2$, HCN, H$_2$CO and H$_2$S have been 
lowered by 2, 6, 2 and 2 respectively, for the sake of clarity. 
The solid lines show the results for the final cycle, 
with results from the previous cycles shown by the dotted lines.}
\label{fig:results_fig3}
\end{figure}

As noted above the chemical effects of the desorption are mostly self-contained
within each cycle and the changes on longer timescales are mostly fairly marginal.
Hence the similarity between the results obtained with the sporadic and the 
continuous desorption formalisms. 
Whilst this is as expected when $t_{\rm impact}$ greatly exceeds the chemical
timescales, this will not be the case if the timescales are comparable (e.g. if
the gas density is lower, or the cosmic ray ionization rate, $\zeta$ is higher).
To illustrate this we have run Model 2 with $\zeta = 5\times 10^{-16}$s$^{-1}$
(for $a=0.1\mu$m, $n=10^5$cm$^{-3}$ and $A_{\rm v}=10$), and 
the results are shown in Figure~\ref{fig:results_fig3} and 
Table~\ref{tab:results_hiz}. 
These are clearly very different to what was obtained with the lower value of 
$\zeta$; the relative abundances are strongly modified (and in fact are closer 
to what was obtained for Model 3). Most importantly, they show that the shorter 
interval between desorption events leads to chemical variations in one cycle 
being carried forward to the next cycle. 
This is quantified in Table~\ref{tab:results_hiz}, which gives the mean abundances 
in each of the five desorption cycles.
From this we can see that, relative to table~\ref{tab:results_loden},
the abundances of most species are enhanced by at least two orders of magnitude 
and, in some cases, significantly more. This is particularly important in the cases
of CO and, especially, H$_2$O where the transient enhancements discussed above are
transmitted from cycle to cycle. 
Significant (and in some cases non-monotonic) cycle-to-cycle variations for most 
species are also evident; whilst the abundances of some species decline  
(e.g. H$_2$O, NH$_3$, O$_2$, HCN, HNC and H$_2$S), the abundances of others rise
(e.g. HCO$^+$, N$_2$ and N$_2$H$^+$). The cycle-dependence is 
particularly dramatic in the cases of NH$_3$ and O$_2$, for which the mean 
abundances both decline by a factor of $\sim$100 from cycle 1 to 5.

Although the cycle duration is much shorter, five cycles (after which cycle-to-cycle
chemical equilibrium has still not been established) corresponds to a time interval
of 0.13\,Myr.
It is therefore quite obvious that the continuous desorption representation breaks 
down in these circumstances and that this situation is one that would be very hard 
to quantify and apply to standard astrochemical models.

\begin{table}
\caption{Results from model 2 with an enhanced cosmic-ray ionization rate
($\zeta = 5\times 10^{-16}$s$^{-1}$): time-averaged fractional abundances 
of selected species for the first 5 desorption cycles.}
\begin{center}
\begin{tabular}{|l|lllll|}
\hline
Species & Cycle 1 & Cycle 2 & Cycle 3 & Cycle 4 & Cycle 5 \\
\hline
CO     & 7.9(-5) & 7.6(-5) & 7.5(-5) & 7.4(-5) & 7.2(-5) \\
H$_2$O & 2.8(-7) & 4.2(-7) & 1.9(-7) & 6.1(-8) & 2.5(-8) \\
NH$_3$ & 2.0(-5) & 8.9(-6) & 2.9(-6) & 6.8(-7) & 2.0(-7) \\
O$_2$  & 1.1(-6) & 8.1(-7) & 3.2(-7) & 6.3(-8) & 1.2(-8) \\
HCO$^+$ & 2.0(-9) & 3.3(-9) & 4.8(-9) & 6.1(-9) & 7.1(-9) \\
H$_2$S & 7.4(-9) & 6.7(-9) & 4.8(-9) & 3.1(-9) & 2.1(-9) \\
HCN    & 1.6(-6) & 3.0(-6) & 2.2(-6) & 1.4(-6) & 9.3(-7) \\
HNC    & 1.3(-6) & 2.4(-6) & 1.8(-6) & 1.1(-6) & 7.2(-7) \\
N$_2$  & 6.7(-6) & 1.1(-5) & 1.5(-5) & 1.8(-5) & 1.9(-5) \\
N$_2$H$^+$ & 9.2(-11) & 2.2(-10) & 3.9(-10) & 5.2(-10) & 6.1(-10) \\
\hline
\end{tabular}
\end{center}
\label{tab:results_hiz}
\end{table}

\section{Comparison with other studies}
\label{sec:compare}

The various issues that we have discussed in this work (extended cooling lifetimes,
a corrected model of the application to interstellar grain size distributions, 
treatment of the desorption process as sporadic, as opposed to continuous, 
the possibility of species co-desorption and whole mantle evaporation from 
very small grains) have not been discussed in any detail in previous studies,
which makes direct comparison difficult.
However it is useful to place this work in the context of these other studies 
and to identify the sources of uncertainty, where known.

There have been several recent studies of cosmic-ray induced desorption in the 
literature. Some studies have addressed the physics of the interaction between 
the cosmic ray particles and the dust grains, e.g. \citet{IRVC15} considered 
a model of localised spot-heating driven explosions,
whilst \citet{KK20a} is the only other investigation that has considered 
the time-dependence of the coupled cooling, and layer-by-layer 
evaporative cooling of different molecular species. 
That study tracked the cooling profile, the composition of 
the ices and the evaporative yields for several species. However, it did not
determine the effective desorption rates or 
consider the chemical effects on an environment that is subject to repeated 
heating and desorption events.
\citet{KK20b} extended the study to include the effects of variations 
in the grain heat capacity and also concluded that the desorption is 
most efficient for small grains, with CO desorption dominated by grains
with $a<0.05\mu$m.

As stated above, for the reasons given, we have conducted our study within, and
extending, the paradigm stated in HH93 - centred on a study of the sporadic
heating of 0.1$\mu$m grains to a peak temperature of $\sim$70K, at the rate of
one event per Myr.
In reality, there will be a spectrum of cosmic rays of varying composition and 
stopping power impacting grains of varying composition and structure.
Furthermore, the relationship between these parameters and the cosmic ray 
energy/flux is not necessarily linear (e.g. see eqn 16 of \citet{LJO85}, 
implying that - for a given grain size - there will be a spectrum of values 
for $T_{\rm peak}$ and $t_{\rm impact}$.
Other studies have therefore concentrated more on taking the standard whole 
grain heating, continuous desorption, model and adapting it for a range of 
grain sizes, ice mantle thicknesses and cosmic ray energy spectra, including any
attenuation effects that may be appropriate.
Thus, \citet{K16} extended the HH93 model to consider the effects of a 
spectrum of cosmic rays 
on grains of bare radius 0.05-0.2$\mu$m and varying ice thickness in a 
translucent cloud ($A_{\rm v}=2$). They found that $t_{\rm impact}$ may 
be over-estimated by two orders or magnitude or more, particularly for 
low-energy impacts resulting in $T_{\rm peak}\sim 20-30$K. This is probably 
too low to effect efficient desorption, but can promote surface chemistry.
\citet{ZCL18} investigated the dependence of the desorption rates on grain 
size and found that the differential depletion of volatile species is
sensitive to the population of large grains ($>0.1\mu$m) that are heated
to lower temperatures, although - as discussed in section~(\ref{sec:MRN}) -
this finding was based on an erroneous definition of the grain cooling timescale.
This was also used by \citet{SZC20} who determined the dust grain equilibrium 
temperatures resulting from CR impact heating; which have important effects 
on the thermal diffusion and surface reaction rates for solid-state species. 
In that study they investigated the dependence of the desorption efficiencies 
on the grain size and found that the desorption rate is weighted towards the 
larger grains in a normal interstellar size distribution.
\citet{KK19} considered the chemical effects of a spectrum of cosmic rays 
resulting in a range of values for $T_{\rm peak}$ and $t_{\rm impact}$. 
The results are presented in the context of a specific physical evolutionary 
model of a contracting dense core.
This study also gives analytical relationships between $T_{\rm peak}$ and 
$t_{\rm impact}$, as well as their dependence on the extinction ($A_{\rm v}$).

\citet{SSC21} emphasised how the desorption efficiency depends on the 
structured content of the different layers of the ices, but made the
assumption that the binding energies are well-defined for each
species and not dependent on their environment. In their model, the 
cooling (but not the temperature-dependences of the desorption rates 
themselves) was treated time-dependently.
A cosmic ray energy/flux spectrum was used to determine $t_{\rm impact}$,
but the complexities of the situation are apparent in that $t_{\rm impact}$
depends on a variety of factors including the shape and strength of the cosmic ray
spectrum, the nature of the particles, the depth-dependence of the 
attenuation of the flux and how that varies as a function of the energy, and hence 
the depth-dependence of the dust opacity and the grain size distribution.
\citet{SCI21} have presented a very complex model which describes desorption as 
a function of the size and composition of both the core grain and the ice mantle,
and includes a cosmic ray transport model with a depth-dependent spectrum. 
They found that there are strong variations of the ice mantle thickness with
grain size.

Taking these various issues together,
this is potentially a very complicated problem to solve; even if the nature and
energy transport of the cosmic rays were known, there are additional uncertainties 
in the microphysics; in the WGH model the rate of thermal diffusion within the 
grains depends on their composition, shape and structure/porosity, whilst the
desorption rates will be moderated by the fact that there will be grain size- and 
source-specific variations in the compositional structure of the ice mantles.

There are clearly a large number of factors and a wide range of the key physical
parameters ($T_{\rm peak}$, $t_{\rm impact}$ etc.) that need to be investigated
further using the results from our study.
However, a key feature of most of these studies is that, by considering the 
spectrum of cosmic rays, heating events (leading to low values of $T_{\rm peak}$)
occur much more frequently than once every 1Myr - possibly by a factor of 
100$\times$ or more. This introduces additional complexity as the lower peak
temperatures will result in increased segregation between the desorption 
efficiencies of volatile and non-volatile species. Most significantly, the 
discrepancies between the sporadic and the continuous desorption approaches 
will be evident and must be considered.

For the purpose of comparison, a useful compromise and simplification is
given by \citet{K21} who identify a single, weighted mean, value of 54K for 
$T_{\rm peak}$ and a simple analytical dependence of $t_{\rm impact}$ on the 
hydrogen column density.
We have tested this in our Model 2 (sporadic desorption) with $A_{\rm v}=10$ 
and density $10^5$cm$^{-3}$ (as for the models of low density regions 
discussed above), for which the implied period between impacts is 
$t_{\rm impact}=10^{11}$s. Other parameters are as given in Table~\ref{tab:params},
as before.

The results, given in the form of the mean cycle-averaged fractional abundances of 
several gas-phase species, are shown in Table~\ref{tab:results_54K}. These 
demonstrate that, for these parameters (and particularly due to the much lower 
value of $t_{\rm impact}$), the behaviours are similar to the case that we 
investigated above for $\zeta =5\times 10^{-16}$\,s$^{-1}$, in that
the chemical evolution of individual desorption cycles are not independent of 
eachother
and the abundances of most species (including CO and H$_2$O) are very 
significantly enhanced.
The table shows how these variations can be carried over after many tens of cycles.
These changes are especially noticeable in the case of NH$_3$, whose mean 
abundance falls by a factor of $>170$ between cycles 20 and 100.
Indeed, these variations are still evident and an equilibrium `limit cycle' has 
still not been established even after 100 cycles (i.e. $\sim$0.3\,Myrs).

\begin{table}
\caption{Results from Model 2 for selected species, with $T_{\rm peak}=54$K 
and $t_{\rm impact}=10^{11}$s. The second column gives the desorption rate 
(s$^{-1}$), and columns 3-6 give the mean (cycle-averaged) fractional abundances 
after 1, 10, 20 and 100 cycles respectively.
For H$_2$O, results are given for two values of the binding energy 
(see section~\ref{sec:extend} for details).}
\begin{center}
\begin{tabular}{|l|l|llll|}
\hline
Species & $k_{54}(0.1\mu$m) & Cycle 1 & Cycle 10 & Cycle 20 & Cycle 100 \\
\hline
CO     & 1.3(-11) & 4.3(-6) & 2.2(-5) & 2.4(-5) & 1.4(-5) \\
CH$_4$ & 6.1(-13) & 1.4(-8) & 7.2(-8) & 1.0(-7) & 1.2(-7) \\
H$_2$O & 2.1(-41) & 8.2(-10) & 8.3(-8) & 2.2(-7) & 2.4(-8) \\
H$_2$O$^{\dag}$ & 2.5(-17) & 8.3(-10) & 8.3(-8) & 2.2(-7) & 2.4(-8) \\
NH$_3$ & 2.1(-10) & 2.6(-5) & 2.8(-5) & 1.7(-5) & 9.6(-8) \\
N$_2$  & 1.3(-11) & 3.4(-7) & 3.7(-6) & 5.4(-6) & 3.6(-6) \\
HNC    & 2.1(-14) & 1.7(-8) & 6.7(-7) & 1.0(-6) & 1.5(-7) \\
H$_2$CO & 1.4(-16) & 2.1(-12) & 8.9(-10) & 3.3(-9) & 1.6(-8) \\
\hline
\end{tabular}
\end{center}
\label{tab:results_54K}
\end{table}

\section{Discussion and conclusions}
\label{sec:summary}

In this study we have extended and developed models of the cosmic ray heating 
induced desorption of species from ice mantles to include:
a description of the grain cooling and desorption profiles, 
a revised analysis of grain size population-averaged rates,
consideration of upper limits to the rates, set by the availabilities 
within the ice mantle,
treatment of the process as being sporadic, rather than continuous,
the possibility of species co-desorption, and
whole mantle desorption from very small grains. 

We do not make any claims to be definitively accurate in quantifying the
cosmic-ray induced desorption rates. However,
this study is based on the energetics of the process; balancing the heating rate 
by the energy lost through the evaporative cooling of the ice constituents.
Because of this, our results are reasonably robust to any assumption
concerning the precise nature of the desorption mechanism and the uncertainties
in the microphysics of the interactions between the cosmic ray particles and 
dust grains.
Using the conservation of energy approach, the only controlling parameters in the 
determination of the desorption rates are 
(i) the mean energy deposited in a grain per cosmic ray impact 
($\Delta E$), 
(ii) the mean interval between cosmic ray impacts ($t_{\rm impact}$), and 
(iii) the binding energies of the atomic and molecular species in 
the dust ice mantles ($E_{\rm B}^{\rm i}$).
Unfortunately, these are all ill-defined so that there are siginificant 
uncertainties in the derived desorption rates.

Compared to previous studies, which are based on a continuous representation of 
the cosmic-ray induced desorption process, 
the two over-arching conclusions of this study are that (a) the cosmic-ray induced 
desorption rates are highly uncertain, and (b) they have almost certainly been 
underestimated - and possibly particularly so for the more tightly bound species, 
such as H$_2$O.

Whatever model we adopt
there are huge uncertainties (possibly by orders of magnitude) due to the 
poor constraints on the many free parameters, some of which were highlighted 
in the previous section. 
It is therefore currently inappropriate to make any specific astrochemical 
deductions.
Indeed, it may therefore be necessary to `invert' the problem and to use 
observationally deduced molecular abundances in dense regions (with reasonably
well-defined physical conditions) to infer the desorption rates.

Specifically, our findings are that:
\begin{enumerate}

\item Cosmic-ray induced desorption rates are almost certainly very much larger 
than the values that have been adopted in most previous studies.
Even within the simple paradigm of HH93 (a single grain size, impacted by cosmic 
rays of the same energy/composition with a single time interval between impacts)
a realistic representation of the cooling/desorption profiles indicates that the 
desorption rates (especially those of the more volatile species) should be
$\gtappeq 10\times$ larger. Generally, the significance of this is only apparent 
at higher densities ($\gtappeq 10^5$cm$^{-3}$), where the role of cosmic-ray
induced desorption is more significant.

\item A consideration of the dependences of the heating rates, peak dust grain 
temperatures and the cooling rates for grains of different sizes shows that, for 
grain size distribution typical of the ISM, desorption from the smallest grains
dominates. This is different to what has found in previous studies, which have 
used a different definition of the grain cooling timescale.
The small grains will achieve higher peak temperatures than the classical
$0.1\mu$m grains so that the most important effect of this is that the desorption 
rates for the more tightly-bound species are significantly enhanced.

\item Desorption rates are limited by the availability of the desorbed species 
within the ice mantles. In practice, this is equivalent to saying that, using the 
HH93 formalism and parameters, $k_{\rm crd}^{\rm i}<10^{-12}$s$^{-1}$.

\item The combined effect of the points highlighted above is to result in a
somewhat compressed dynamic range for $k_{\rm crd}^{\rm i}$.
Recognising that the binding energies of molecular species in ices are
highly uncertain (especially in the case of mixed composition ice) we have 
provided a table of the desorption rates (for both the `classical' 0.1$\mu$m 
grains and for an MRN-weighted distribution of grain sizes) as a function
of binding energy. This data can easily be incorporated into astrochemical
models. 

\item If we do {\em not} make the usual assumption that the sporadic/stochastic
mechanism of cosmic ray impacts/heating/desorption can be represented as a 
continuous but instead follow the chemical evolution through sucessive heating 
and desorption cycles; we obtain similar results for most species (but not all; 
most notably there are large differences for CO and NH$_3$) - but {\em only} 
if the desorption cycles are chemically independent of eachother.
However, the discrepancies between the sporadic and the continuous 
representations become significant when the interval between cosmic ray impacts
is shorter than the chemical timescales.
The key timescale here is that for freeze-out so that, using the HH93 
description for $0.1\mu$m grains, significant cycle-to-cycle
chemical variations occur when
\[ \zeta > 1.3\times 10^{-17}s^{-1} (n/2\times 10^4 cm^{-3}). \]
More generally, this is when 
\[ t_{\rm impact} < 10^{11}s. (6\times 10^6 cm^{-3}/n). \]
In these circumstances, which are not expected to be unusual in star-forming 
cores,
two results follow: (a) the abundances (e.g. of H$_2$O and CO) may be very 
significantly enhanced with highly non-linear $\zeta$-dependences, and (b)
the chemical evolution of many species (e.g. NH$_3$)
will occur over many cycles and the timescale for
the changes will be significantly larger than that for either the 
freeze-out of gas-phase species, or the interval between cosmic-ray impacts.
Obviously, in these conditions, the continuous desorption simplification 
is not applicable.
Recent studies, which include a spectrum of cosmic ray energies and therefore a
range of values of $t_{\rm impact}$, indicate that these effects cannot be 
ignored.

\item Whether or not different ice components can co-desorb - where the 
desorption characteristics of one species affects others - is somewhat 
speculative and will be strongly dependent on the compositinal structure of the 
ices. However,
the presence of even limited levels of species co-desorption results in very large
($\gtappeq 10-100\times$) enhancements of the effective desorption rates and 
predicted gas-phase abundances (especially for densities $\gtappeq 10^5$cm$^{-3}$).

\item Similar, but more species-specific, enhancements are obtained if
a population of very small grains exist, from which complete ice mantle
sublimation is possible following each cosmic ray heating event. Here, the 
strongest enhancements are for the more tightly bound species, so that a 
chemical signature of the presence of these grains may be detectable.
Generally, if either co-desorption occurs and/or very small grains can 
survive to high densities, then the desorption rates from tightly bound 
species (such as H$_2$O) would be very strongly enhanced.

\item
%
At extinctions $A_{\rm v}\gtappeq 5.5$, which applies throughout most 
of the gas in dense star-forming cores, photodesorption is dominated by 
the cosmic-ray induced radiation field. The photodesorption rates 
(cm$^{-3}$s$^{-1}$) are given by
$ \dot{n}_{\rm i}\, = Y_{\rm i} F_{\rm cr} \sigma_{\rm H} n_{\rm H} f_{\rm i}$
where $Y_{\rm i}\sim 10^{-3}$ is the photodesorption yield and $F_{\rm cr}$ 
is the cosmic-ray induced photon flux (typically $\sim 5,000$ 
photons cm$^{-2}$\,s$^{-1}$).
Comparing this to the desorption rate for cosmic-ray heating (equation \ref{eqn:crdes}), 
we can see that (noting that both processes scale with $\zeta$) desorption by 
cosmic ray heating dominates over photodesorption when 
\[ k_{\rm crd}^{\rm i} > \frac{Y_{\rm i}F_{\rm cr}}{N_{\rm s}} 
\sim 5\times 10^{-15} s^{-1}. \]
Comparing this to the values given in Tables~\ref{tab:rates_HH93},
\ref{tab:rates_1200} and \ref{tab:rates_1100} we can conclude that (even without
the additional enhancements to the rates due to sporadic heating, co-desorption 
or the presence of small grains) cosmic ray heating will dominate over 
photodesorption for most species in dense, dark environments.
This is an important result that is in direct contrast to the findings of 
previous studies \citep[e.g.][]{SGSD04} and is a consequence of the higher 
values that we have determined for the cosmic-ray induced desorption rates.
To illustrate this point, we have re-run several of the models including the full
range of desorption mechanisms (direct and cosmic ray-induced photodesorption, 
enthalpy of formation/H$_2$ formation driven desorption etc.) and a limited surface
chemistry. This model is therefore closer to that which might be used to simulate 
real molecular clouds/star-forming regions.
The results, for selected species, are shown in Table~\ref{tab:results_alldes}.
This table also includes results from a reference model (as model 1, but without any
cosmic ray desorption) for comparison.

\begin{table}
\caption{Time-averaged fractional abundances of selected species for
models in which all desorption mechanisms are included 
($n=10^7$cm$^{-3}$, $A_{\rm v}$=50). See text for details.} 
\begin{center}
\begin{tabular}{|l|llll|}
\hline
Species & Reference & Model 1c & Model 2 & Model 3a \\
\hline
CO     & 8.9(-11) & 1.7(-9) & 2.3(-9) & 2.1(-9) \\
H$_2$O & 3.2(-11) & 3.4(-11) & 3.2(-11) & 2.0(-9) \\
NH$_3$ & 6.0(-13) & 1.1(-9) & 7.3(-9) & 4.9(-11) \\
HCO$^+$ & 8.6(-14) & 1.6(-12) & 1.1(-13) & 1.2(-13) \\
CO$_2$ & 7.4(-12) & 8.2(-12) & 3.2(-12) & 1.1(-10) \\
HCN    & 2.6(-15) & 9.1(-15) & 1.4(-14) & 1.3(-13) \\
N$_2$  & 1.2(-11) & 1.9(-10) & 1.5(-10) & 2.9(-10) \\
\hline
\end{tabular}
\end{center}
\label{tab:results_alldes}
\end{table}

These results clearly show the significance of cosmic ray induced desorption,
with the abundances of many species showing dramatic differences relative to 
the reference model. Whilst the effects on more tightly bound species, such as 
H$_2$O and CO$_2$, are more marginal these are also dramatically enhanced if 
co-desorption occurs.

\end{enumerate} 

The significance of these results is most apparent in the dense, 
dark regions whose chemistry and dynamics are used to diagnose the physics
of the early stages of star-formation, and which may form the chemical
`legacy' to protoplanetary disks.

Finally, we note that
there are additional implications of the cosmic-ray interactions that we 
have not considered in this study, such as the effects on the grain charge
\citep{IPGC15} or the cosmic-ray driven non-thermal chemistry within the 
ices \citep{SCI21}.
In future work, we will assess the importance of the updates that we have 
discussed in this study in realistic astrochemical models of specific 
sources. To do that we will also consider the application to a range of grain 
sizes and (possibly attenuated) cosmic ray energy spectra, whilst being mindful 
of the many uncertainties that exist.

\section*{Acknowledgement}

The author thanks the anonymous referee for helpful comments and 
suggestions that have improved this paper.

\section*{Data availability}

The data underlying this study are openly available from the published
papers that are cited in the article.
No new data were generated in support of this research.

\bsp
\label{lastpage}

\begin{thebibliography}{}

\bibitem[Bringa \& Johnson(2004)]{BJ04}
Bringa E.M., Johnson R.E., 2004, ApJ, 603, 159

\bibitem[Bringa \& Johnson(2000)]{BJ00}
Bringa E.M., Johnson R.E., 2000, Surf. Sci., 451, 108

\bibitem[Caselli et al.(2012)]{Cas12}
Caselli P., Keto E., Bergin E.A., Tafalla M., Aikawa Y.,
Douglas T., Pagani L., Y1ld1z U.A., van der Tak F.F.S.,
Walmsley C.M., Codella C., Nisini B., Kristensen L.E., 
van Dishoeck E.F., 2012, ApJ, 759, L37

\bibitem[Cecchi\textendash Pestellini et al.(2010)]{CCP10} 
Cecchi\textendash Pestellini C., Rawlings J.M.C., Viti S., Williams D.A., 
2010, ApJ, 725, 1581

\bibitem[Cernicharo et al.(2012)]{Cern12}
Cernicharo J., Marcelino J., Roueff E., Gerin M., 
Jim\'{e}nez-Escobar A., Mu\~{n}oz Caro G.M., 2012, ApJ, 759, L43

\bibitem[Collings et al.(2004)]{Cetal04} 
Collings M.P., Anderson M.A., Chen R. et al., 2004,
MNRAS, 354, 1133

\bibitem[Collings et al.(2015)]{Cetal15} 
Collings M.P., Frankland V., Lasne J., Marchione D., Rsu-Finsen A., 
McCoustra M.R.S., 2015, MNRAS, 449, 1826

\bibitem[d'Hendecourt et al.(1982)]{HABG82}
d'Hendecourt L.B., Allamandola L.J., Baas F., Greenberg J.M., 1982,
A\&A, 109, L12

\bibitem[Hasegawa \& Herbst(1993)]{HH93}
Hasegawa T.I., Herbst E., 1993, MNRAS, 261, 83

\bibitem[Herbst \& Cuppen(2006)]{HC06}
Herbst E., Cuppen H.M., 2006, Proceedings of the National Academy of 
Science, 103, 12257

\bibitem[Ivlev et al.(2015a)]{IRVC15}
Ivlev A.V., R\"{o}cker T.B., Vasyunin A., Caselli P., 2015, ApJ, 805, 59

\bibitem[Ivlev et al.(2015b)]{IPGC15}
Ivlev A.V., Padovani M., Galli D., Caselli P., 2015, ApJ, 812, 135

\bibitem[Kalv$\overline{\mbox{a}}$ns(2016)]{K16}
Kalv$\overline{\mbox{a}}$ns J., 2016, ApJS, 224, 42

\bibitem[Kalv$\overline{\mbox{a}}$ns(2021)]{K21}
Kalv$\overline{\mbox{a}}$ns J., 2021, ApJ, 910, 54

\bibitem[Kalv$\overline{\mbox{a}}$ns \& Kalnin(2019)]{KK19}
Kalv$\overline{\mbox{a}}$ns J., Kalnin J.R., 2019, MNRAS, 486, 2050

\bibitem[Kalv$\overline{\mbox{a}}$ns \& Kalnin(2020a)]{KK20a}
Kalv$\overline{\mbox{a}}$ns J., Kalnin J.R., 2020a, A\&A, 633, A97

\bibitem[Kalv$\overline{\mbox{a}}$ns \& Kalnin(2020b)]{KK20b}
Kalv$\overline{\mbox{a}}$ns J., Kalnin J.R., 2020b, A\&A, 641, A49

\bibitem[L\'{e}ger, Jura \& Omont(1985)]{LJO85} 
L\'{e}ger A., Jura M., Omont A., 1985, A\&A, 144, 147

\bibitem[Mathis, Rumpl \& Nordsieck(1977)]{MRN77}
Mathis J.S., Rumpl W., Nordsieck K.H., 1977, ApJ, 217, 425

\bibitem[Minissale et al.(2016)]{Min16}
Minissale M., Dulieu F., Cazaux S., Hocuk S., 2016, A\&A, 585, A24

\bibitem[\"{O}berg et al.(2005)]{Ob05}
\"{O}berg K. I., van Broekhuizen F., Fraser H.J., Bisschop S.E., 
van Dishoeck E.F., Schlemmer S., 
2005, ApJ, 621, L33

\bibitem[\"{O}berg, van Dishoeck \& Linnartz(2009)]{ODL09}
\"{O}berg K.I., van Dishoeck E.F., Linnartz H., 
2009, A\&A, 496, 281

\bibitem[\"{O}berg et al.(2009)]{Ob09}
\"{O}berg K.I., Linnartz H., Visser R., van Dishoeck E.F., 
2009, ApJ, 693, 1209

\bibitem[Prasad \& Tarafdar(1983)]{PT83}
Prasad S.S., Tarafdar S.P., 1983, ApJ, 267, 603

\bibitem[Rawlings et al.(2013)]{Raw13}
Rawlings J.M.C., Williams D.A., Viti S., Cecchi-Pestellini C.,
Duley W.W., 2013, MNRAS, 430, 264 

\bibitem[Rawlings \& Williams(2021)]{RW21}
Rawlings J.M.C., Williams D.A., 2021, MNRAS, 500, 5117 


\bibitem[Roberts et al.(2007)]{RRVW07}
Roberts J.F., Rawlings J.M.C., Viti S., Williams D.A., 2007,
MNRAS, 382, 733

\bibitem[Schutte \& Greengerg(1991)]{SG91}
Schutte W.A., Greenberg J.M., 1991, A\&A, 244, 190

\bibitem[Shen et al.(2004)]{SGSD04}
Shen C.J., Greenberg J.M., Schutte W.A., van Dishoeck E.F.,
2004, A\&A, 415, 203

\bibitem[Shingledecker et al.(2018)]{STGH18}
Shingledecker C.N., Tennis J., Le Gal R., Herbst E., 
2018, ApJ, 861, 20

\bibitem[Silsbee, Caselli \& Ivlev(2021)]{SCI21}
Silsbee K., Caselli P., Ivlev A.V., 2021, MNRAS, 507, 6205

\bibitem[Sipil\"{a}, Zhao \& Caselli(2020)]{SZC20}
Sipil\"{a} O., Zhao B., Caselli P., 2020, A\&A, 640, A94

\bibitem[Sipil\"{a}, Silsbee \& Caselli(2021)]{SSC21}
Sipil\"{a} O., Silsbee K., Caselli P., 2021, ApJ, 922, 126

\bibitem[Tafalla et al.(2004)]{TMCW04}
Tafalla M., Myers P.C., Caselli P., Walmsley C.M.,
2004, A\&A, 416, 191

\bibitem[van Dishoeck et al.(2021)]{vD21}
van Dishoeck E.F. et al., 
2021, A\&A, 648, 24

\bibitem[Viti et al.(2004)]{Vetal04} 
Viti S., Collings M.P., Dever J.W., McCoustra M.R.S., Williams D.A., 2004,
MNRAS, 354, 1141

\bibitem[Watson(1976)]{Wat76}
Watson W.D., 1976, Rev. Mod. Phys., 48, 513

\bibitem[Zhao, Caselli \& Li(2018)]{ZCL18}
Zhao B., Caselli P., Li Z-Y., 2018, MNRAS, 478, 2723

\end{thebibliography}
\end{document}